\documentclass[useAMS,usenatbib,a4paper]{mn2e}
\usepackage{amssymb,amsmath}
\usepackage{graphicx}
\usepackage{natbib}
\usepackage{journal_shortcuts}

\newcommand{\be}{\begin{equation}}  \newcommand{\ba}{\begin{eqnarray}}
\newcommand{\ee}{\end{equation}}  \newcommand{\ea}{\end{eqnarray}}
\newcommand{\Bubble}{_{\mathrm{bbl}}}

\newcommand{\Inj}{_{\mathrm{inj}}}

\title[Evolution of X-ray cavities]{Evolution of X-ray cavities}

\author[M. Br\"uggen]  
{Marcus Br\"uggen$^1$\footnotemark[1], Evan
Scannapieco$^2$ and Sebastian Heinz$^3$ \\ $^1$Jacobs University
Bremen, P.O. Box 750\,561, 28725 Bremen, Germany\\ $^2$School of Earth
and Space Exploration,  Arizona State University, P.O.  Box 871404,
Tempe, AZ, 85287-1404, USA\\ $^3$Department of Astronomy, University
of Wisconsin, 475 N Charter Street  Madison, WI 53706, USA }

\begin{document}

\date{Accepted. Received; in original form }

\pagerange{\pageref{firstpage}--\pageref{lastpage}} \pubyear{2007}

\maketitle

\label{firstpage}

\begin{abstract}

A wide range of recent observations have shown that AGN-driven
cavities may provide the energy source that balances the cooling
observed in the centres of ``cool core'' galaxy clusters.  One tool
for better understanding the physics of these cavities is their
observed morphological evolution, which is dependent on such
poorly-understood properties as the turbulent density field and the
impact of magnetic fields. Here we combine numerical simulations that
include subgrid turbulence and software that produces synthetic X-ray
observations to examine  the evolution of X-ray cavities in the
absence of magnetic fields. Our results reveal  an anisotropic size
evolution of  that is dramatically different from simplified,
analytical  predictions.  These differences highlight some of the key
issues that must be accurately  quantified when studying AGN-driven
cavities, and help to explain why the inferred $pV$ energy  in these
regions appears to be correlated with their distance from the cluster
center.  Interpreting X-ray observations  will require detailed
modeling of effects including mass-entrainment, distortion by drag
forces, and  projection. Current limitations do not allow a
discrimination between purely hydrodynamic and magnetically-dominated
models for X-ray cavities.

\end{abstract}


%*******************
\section{Introduction}
%*******************

Feedback from active galactic nuclei (AGN) is widely believed to be
the most promising mechanism to solve a number of problems in our
understanding of the history of baryonic structure formation,
including:  (i) regulating  supermassive black hole formation
\citep{churazov:05,sijacki:06}; (ii) setting the upper mass cut-off in
the galaxy mass function, and determining its evolution with
redshift \citep{benson:03, best:05};   (iii) providing the
nongravitational ``preheating" that is likely to have been important
in determining the entropy profiles of non ``cool-core'' galaxy clusters
\citep{donahue:06, scannapieco:04}; and  (iv) offsetting  the
radiative losses seen in cool-core galaxy clusters
\citep[see e.g.,][]{mcnamara:07}.  Such clusters show strong peaks in their
central X-ray surface brightness distributions, indicating that the
central gas is cooling rapidly.  Yet the deficit of star formation and
$<$1 keV gas \citep[e.g.,][]{rafferty:06} means that radiative cooling
must be balanced by an unknown energy source. Currently, the most
successful model for achieving this balance relies on heating from a
central AGN; yet the details of this process are poorly understood
\citep[e.g.,][]{bruggen:02, reynolds:02, brighenti:06}.

While AGN in cool-core clusters are observed to drive large bubbles
into the intracluster medium (ICM)
\citep{mcnamara:00, blanton:01, finoguenov:01, nulsen:05}, the
synchrotron radiation emitted by the relativistic electrons in these bubbles fades
and becomes extremely difficult to detect after about $10^8$ years. 
Moreover, the corresponding  X-ray surface brightness depressions 
are only visible near the centre of the cluster where the
contrast is large. AGN have also been observed to induce shocks and/or
sonic motions in the ICM that are believed to dissipate
their energy into this gas \citep{fabian:03, fabian05, kraft:07,
ruszkowski04, mcnamara:05, finoguenov:08,sanders:08}, although the
impact of the resulting heating is difficult to quantify
observationally.  Thus, it is unclear how far AGN-driven cavities rise in
the cluster, how they couple to the surrounding medium, and how they
evolve.

More specifically,  the presence of these cavities 
has raised a number of key questions.  The buoyant bubbles inflated by
the central AGN are unstable  to the Rayleigh-Taylor (RT) instability,
which occurs whenever a fluid  is accelerated or supported against
gravity by a fluid of lower density. Yet, these cavities appear to be
intact even after inferred ages of several $10^8$ yrs, as is the case for
the outer cavities in Perseus \citep{nulsen:05}. On the other hand, purely hydrodynamic
simulations fail to reproduce these observations.  Instead, the RT and other
instabilities shred the bubbles in a relatively short time
\citep{brueggen05, heinz:06, pizzolato:06}. Magnetic fields have been
shown to alleviate this problem somewhat \citep{robinson:04, jones:05,
ruszkowski:07}, but they also reduce the extent to which the interior
of the hot bubbles couples to the surrounding medium, making it much
more difficult for AGN heating to balance cooling.

In a recent paper, Scannapieco \& Br\"uggen (2008) showed that
although pure-hydro simulations indicate that AGN bubbles are
disrupted into pockets of underdense gas, more detailed modeling of
turbulence indicates that this is a poor approximation to a cascade of
structures that continues far below current resolution limits.  Using
a subgrid turbulence model developed by Dimonte \& Tipton (2006), they
carried out a series of simulations of AGN heating in a cool-core
cluster with the adaptive mesh refinement code, FLASH.  These
simulations showed that Rayleigh-Taylor instabilities act on subgrid
scales to effectively mix the heated AGN bubbles with the ICM, while
at the same time preserving them as coherent structures. The AGN
bubbles are thus transformed into hot clouds of mixed material as they
move outwards in the hydrostatic medium, much as large airbursts lead to
a distinctive ``mushroom cloud'' structure as they rise in the
hydrostatic atmosphere of Earth. This allows X-ray cavities to remain
intact out to large distances from the cluster centre while still
coupling  to the surrounding medium.

Alternatively, it has been suggested that instead of underdense
pockets of ideal gas, the cavities are produced by magnetically
dominated jets \citep{li:06}. In 3D MHD simulations by
\cite{nakamura:06, nakamura:07} bubbles are inflated by a
current-carrying jet that injects magnetic flux into a small volume in
the vicinity of the supermassive black hole.  The jets are launched by
injecting non-force-free poloidal and toroidal magnetic fields, and
form large currents which travel along the inner jet axis into the
lobes and return on the outer boundary of the lobes, forming a sheath
around the jet axis.  Like bubbles of hot, underdense gas, such jets
expand  subsonically into wide lobes that appear cooler than the surrounding medium.
Unlike hot, underdense gas, these magnetically dominated bubbles behave differently as
they rise through the cluster atmosphere.

Recently, \cite{diehl:08} investigated whether one can use the
measured sizes of X-ray cavities observed at different locations in
their host clusters to discriminate between these two models.
Compiling the sizes and radial offsets of 64 cavities in the X-ray
halos of clusters and groups, they were able to show a tight
correlation between these two quantities, which is substantially
different than one would expect from simple analytic estimates in the
pure-hydro case.  From this comparison, they  came to the preliminary
conclusion that the data favor the current-dominated
magneto-hydrodynamic jet model.

While a useful first step, such analytical prescriptions neglect some
important physical effects that are likely to affect the
interpretation of the data. While the loss of pressure due 
to bremsstrahlung radiation is negligible, mass entrainment produced by
hydrodynamic instabilities is likely to be important
\citep{pavlovski:07}. This will add to the growth of the
bubbles produced by the expansion during the ascent in the stratified
atmosphere. In fact, there is some circumstantial evidence for mass entrainment
in FR I radio sources.  \cite{croston:08} showed that equipartition
internal pressures are typically lower than the external pressures
acting on the radio lobes, so that additional non-radiating particles
must be present. A correlation between the structure of the radio
sources and the apparent pressure imbalance can be taken as
observational evidence that entrainment may provide this missing
pressure.  Moreover, drag forces and acceleration by buoyancy 
distort the bubbles, complicating the interpretation of bubble radii.

Related to these issues is the question as to how to best measure the
energetics of AGN-inflated bubbles.  It is now standard
procedure to use the size of a bubble and the ambient pressure at its
location to get an estimate of the $pdV$ work needed to inflate the
bubble, which is then used to infer the energy associated with the
outburst \citep[e.g.,][]{birzan:04}. Interestingly, if one plots the
inferred energy as a function of radius for the known X-ray cavities,
one finds that the energy increases as a function of bubble distance
from the cluster centre  \citep[][see Fig.~\ref{fig:pvplot}]{rafferty:06}. 
This effect is not only apparent in a large
sample of different clusters, but even within the few individual
clusters with known multiple bubbles, such as Hydra~A and
Perseus. Recently, \cite{sanders:07} reported the possible detection
of another cavity in Perseus at very large radii ($\approx 170$ kpc), and
inferred an outburst energy several times higher than that inferred
for bubbles at $\approx$ 30 kpc from the cluster centre. Even
larger differences between inner and outer bubbles have been reported
in Abell 2204 where the outer bubbles have radii of 240 kpc and 160
kpc \citep{sanders:08b}. Understanding this effect will be crucial in
correctly addressing the question whether AGN can be a general
solution to the cooling flow problem in clusters and galaxies.

Here we combine adaptive-mesh refinement (AMR) simulations that
include a subgrid turbulence model \citep{scannapieco:08} with
software to produce synthetic X-ray observations to better reproduce
the detailed evolution of hydrodynamic bubbles \citep{heinz:09}.  Using
these results, we are able to  quantify the extent to which
hydrodynamic bubbles obey simple analytic relations and the effect of
subgrid turbulence.  In particular, the inferred sizes of bubbles as a
function of the cluster-centric distance in our simulations allow us
to directly address  the extent to which such observations can be used
to constrain the role of magnetic fields in bubble evolution and
measure the overall energy input from AGN in the centres of cool-core
clusters.

The structure of this works is as follows.  In \S 2 we describe our
simulations, spectral modeling, and analysis methods.   In \S3 we
present our results and compare them with analytical estimates and
observed trends.  Conclusions are given in \S4.

%*************
\section{Method}
%*************

\subsection{Code} \label{sec:code}

All simulations were performed with  FLASH version 3.0 \citep{fryxell:00}, a multidimensional adaptive mesh refinement hydrodynamics
code, which  solves the Riemann problem on a Cartesian grid using a
directionally-split  Piecewise-Parabolic Method (PPM) solver.  While
the direct simulation of turbulence is extremely challenging,
computationally expensive, and dependent on resolution  
\citep[e.g.,][]{glimm:01}, its behavior can be approximated to a good degree of
accuracy by adopting a subgrid approach.

Recently, \cite{dimonte:06}, described a subgrid model that is
especially suited to capturing the buoyancy-driven turbulent evolution
of AGN bubbles. The model captures the self-similar growth of the RT
and Richtmyer-Meshkov (RM) instabilities by augmenting the mean hydrodynamics equations
with evolution equations for the turbulent kinetic energy per unit
mass and the scale length of the dominant eddies. The
equations are based on buoyancy-drag models for RT and RM flows, but
constructed with local parameters so that they can be applied to
multi-dimensional flows with multiple materials.  The model is
self-similar, conserves energy, preserves Galilean invariance, and
works in the presence of shocks. Although it contains several
unknown coefficients, these are determined by comparisons with
analytic solutions, numerical simulations, and experiments.
In Scannapieco \& Br\" uggen (2008), we describe our implementation
of this model into the  FLASH code, present several comparisons with analytic
results, and discuss its usefulness for properly capturing the physics of
AGN-driven cavities in the absence of magnetic fields.  The reader is referred to this work.
for these and other related  discussions.

In the present study,
all our simulations are performed in a cubic three-dimensional region
680 kpc on a side, with all reflecting boundaries.  For our grid, we
chose a block size of $8^3$ zones and an  unrefined root grid with
$8^3$ blocks, for a native resolution of 10.6 kpc.   The refinement
criteria are the standard density and pressure criteria, and we allow
for 4  levels of refinement beyond the base grid, corresponding to  a
minimum cell size of 0.66 kpc, and an effective grid of 1024$^3$
zones.

\subsection{Cluster Profile}

For our overall cluster profile, we adopted the model described in
\cite{roediger:07}, which was constructed to reproduce the properties of the
brightest X-ray cluster A426 (Perseus), which  has been studied
extensively with {\sc Chandra} and {\sc XMM}-Newton.  In this case,
the electron density $n_{\rm e}$  and temperature $T_{\rm e}$  profiles
are based on the deprojected XMM-Newton data \citep{churazov:03} which are also 
in broad agreement  with the  {\sc Chandra} data \citep{schmidt:02, sanders:04}. Namely: 
\be 
n_{\rm
e}=\frac{4.6\times10^{-2}}{[1+(\frac{r}{57})^2]^{1.8}}+
\frac{4.8\times10^{-3}}{[1+(\frac{r}{200})^2]^{0.87}}~~~{\rm cm}^{-3},
\label{eq:ne}
\ee 
and 
\be T_{\rm e}=7\times\frac{[1+(\frac{r}{71})^3]}
{[2.3+(\frac{r}{71})^3]}~~~{\rm keV},
\label{eq:te}
\ee 
where $r$ is measured in kpc. Furthermore, the hydrogen number
density was  assumed to be related to the electron number density as
$n_{\rm H}=n_{\rm e}/1.2$ according to standard cosmic abundances.  The static, spherically-symmetric
gravitational potential was set such that  the cluster was in hydrostatic equilibrium.

\subsection{Bubble generation} \label{sec:bubble_generation}

X-ray cavities in the ICM are thought to be inflated by a pair of ambipolar
jets from an AGN in the central galaxy that inject energy into small
regions at their terminal points, which expand until they reach
pressure equilibrium with the surrounding ICM
\citep{blandford:74}. The result is a pair of underdense, hot bubbles
on opposite sides of the cluster centre.

In order to produce bubbles, we started the simulation by injecting
a total energy of $E\Inj$
into two small spheres of radius $r\Inj$ = 4.5 kpc at distances of 13 kpc from
the cluster centre. The gas inside these spheres was heated and
expanded similar to a Sedov explosion to form a pair of bubbles in a
few Myrs, a time much shorter than the rise time of the generated
bubbles.  The parameters $r\Inj$, and $E\Inj$ were chosen
such that these regions reached a radius of $r\Bubble = 12$ kpc and a
density contrast of approximately  $\rho_{\rm b}/\rho_{\rm amb} =
0.05$ as compared to the surrounding ICM.  
However, the dependence
of the bubble dynamics on the density contrast, $\rho_{\rm
b}/\rho_{\rm amb}$, is weak provided that $\rho_{\rm b}/\rho_{\rm amb}
\ll 1$.
For a $\gamma = 5/3$ gas, the energy
input during this expansion can be simply  calculated as 
\be 
E_{\rm expand} = \int_{V_{\rm init}}^{V_{\rm evac}} dV p
%= p_{\rm final}  \int_{V_{\rm init}}^{V_{\rm final}} dV 
%   \left( \frac{V_{\rm final}}{V_{\rm init} \right)^{5/3}
= \frac{3 p_{\rm evac} V_{\rm evac}}{2}  \left[ \left( \frac{V_{\rm
  evac}}{V_{\rm init}} \right)^{2/3} -1 \right],
\label{eq:expand}
\ee where $p_{\rm evac}$ and $V_{\rm evac}$ are the pressure and
volume at the evacuated stage at the end of the expansion and $V_{\rm init}$
is the initial volume.
Finally, the energy released as the bubble moves outwards to the edge of the cluster is
\be 
E_{\rm bou} = \int_{V_{\rm init}}^{V_{\rm evac}} dV p
%= p_{\rm final}  \int_{V_{\rm init}}^{V_{\rm final}} dV 
%   \left( \frac{V_{\rm final}}{V_{\rm init} \right)^{5/3}
= \frac{3 p_{\rm evac} V_{\rm evac}}{2}.
\label{eq:bou}
\ee
The sum of these two contributions for both bubbles amounted to $1.1 \times 10^{60}$ ergs.
%=(4*3.1415/3.)*(12.*3.1E21)^3*5.E-2*1.6E-9*7./2.3*3./2.*(20.)^(2./3)*2.

\subsection{Synthetic Chandra observations and spectral modeling} \label{chandra}

In order to  allow a direct comparison of
simulation output with X-ray data, we made use of a newly-developed
pipeline for post-processing of gridded simulation output.
The X-ray-imaging pipeline {\tt XIM} \citep[see also][]{heinz:09} is
a publically-available set of scripts that automate the creation of
simulated X-ray data for a range of satellites.  It takes as input the
density, temperature, and velocity, as well as a large number of
parameters and provides simulated X-ray data in the form of spectral-imaging 
data cubes.  Simple batch-processing and trivial parallelisation allow the
manipulation of these large data cubes, provided that the final output data
fit entirely in memory.

{\tt XIM} is focused on visualizing X-ray data from thermal plasmas.
The scripts allow the choice of a user-supplied spectral model as well
as the default thermal {\tt APEC} plasma emission model
\citep{smith:01}, which self-consistently calculates the equilibrium
ionization balance for a thermal plasma.  For a fixed set of
abundances and a given temperature, {\tt APEC} then computes
interpolated high resolution model X-ray spectra.  Atomic data are
taken from the {\tt ATOMDB} using {\tt APED} \citep{smith:01b} and
combined with bremsstrahlung continuum for all species.

Querying {\tt APED} for every simulated cell is computationally too
expensive.  For computational convenience and speed, {\tt XIM} creates
a table of model spectra that span a user-supplied energy vector and
a logarithmic range of temperatures with a user-supplied minimum and
maximum.  The spectral contribution of each cell is then
logarithmically interpolated from this table. In this paper, we have
fixed the  relative  abundance of all elements heavier than helium at
the solar value, but  in principle, the metallicity $Z$ of the gas can
be specified on a cell-by-cell  level, for fixed relative abundances
of heavy elements.

\subsection{Spectral projection}

Having computed the spectral contribution in each cell,
{\tt XIM} then calculated a raw spectral data cube by projecting the
data along one of the three Cartesian coordinate axes.  Spectra were
emission-measure weighted and Doppler shifted with the user-provided
radial velocity, neglecting relativistic effects.  A user-supplied
tracer grid was used to weigh the data by the thermal plasma content
of the cell.  The spectral grid was oversampled by a factor of three with respect to
the final output energy grid to allow accurate representation
of Doppler shifts in the output spectra.

The data was further redshifted according to the user-specified
cosmological redshift and the coordinate axes are scaled to the proper
angular size, given the redshift and cosmological parameters, by
default using concordance parameters $\Omega=1$,
$\Omega_{\Lambda}=0.7$, $H_{0}=70\,{\rm km\,s^{-1}\,Mpc^{-1}}$.  
The flux in a cube (x,y,wavelength) was then scaled to the cosmologically correct 
flux at the given distance.
The projected data were processed for foreground photo-electric
absorption using the Wisconsin Absorption Model \citep[WABS][]{morrison:83}.

\subsection{Virtual observation}

Next,
the raw spectral-imaging cube (x,y,wavelength) was re-gridded in the
two spatial directions onto the detector plate scale of the
user-specified instrument.  {\tt XIM} incorporates telescope
parameters for  {\sc Chandra, Constellation-X, XEUS}, and  {\sc
XMM-Newton} and will incorporate a telescope model for IXO once
response matrices become available, 
in our case the Advanced 
Imaging Spectrometer (ACIS), on board the {\sc Chandra} X-ray observatory.
The re-gridded data cube was then
convolved with the appropriate spectral response and ancillary response matrices,
and the convolution output was regridded
onto the user-specified energy grid.

The output was convolved with a model point-spread function for the selected telescope.
The current version of {\tt XIM} is limited to a Gaussian point spread function with
energy-independent kernel width. It takes into account quantum efficiency and
telescope effective area, but neglects detector
non-uniformity, vignetting, and point-spread function variance.  The latter effects can
in principle be modeled through post-processing with third-party
telescope simulators (e.g., {\tt MARX}) \citep{wise:97}. Finally, Poisson-distributed
photon counts were calculated for a user-specified exposure time.

%FFFFFFFFFFFFFFFFFF
\begin{figure*}
\includegraphics[trim=0 0 0 0,clip,width=0.45\textwidth]{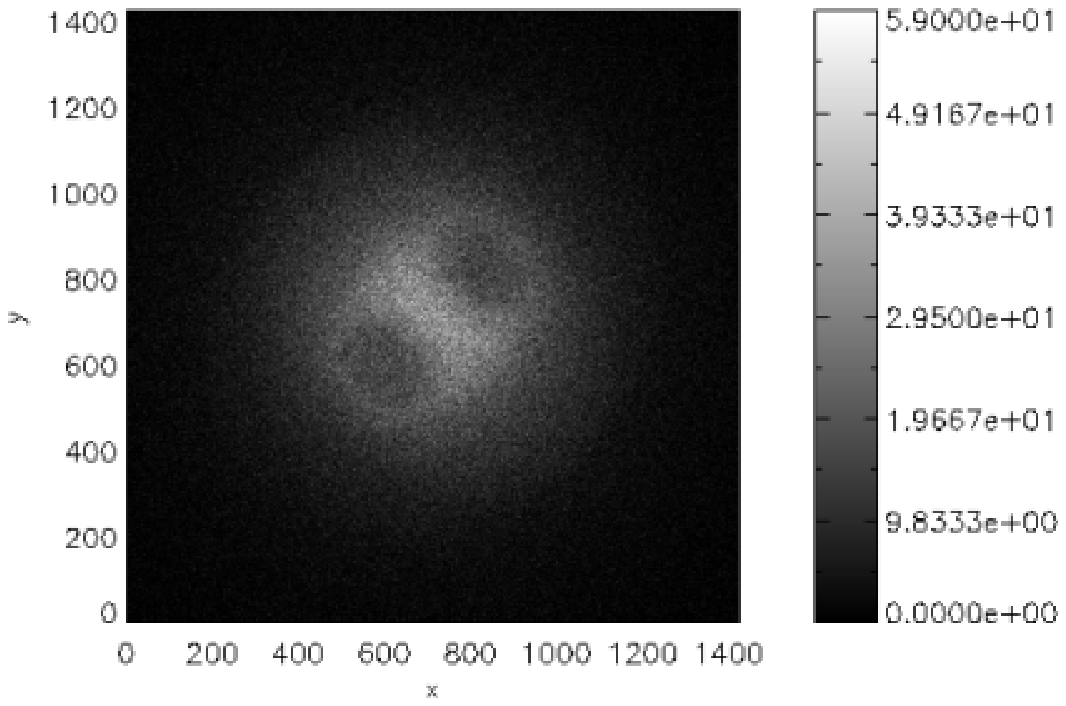}
\includegraphics[trim=0 0 0 0,clip,width=0.45\textwidth]{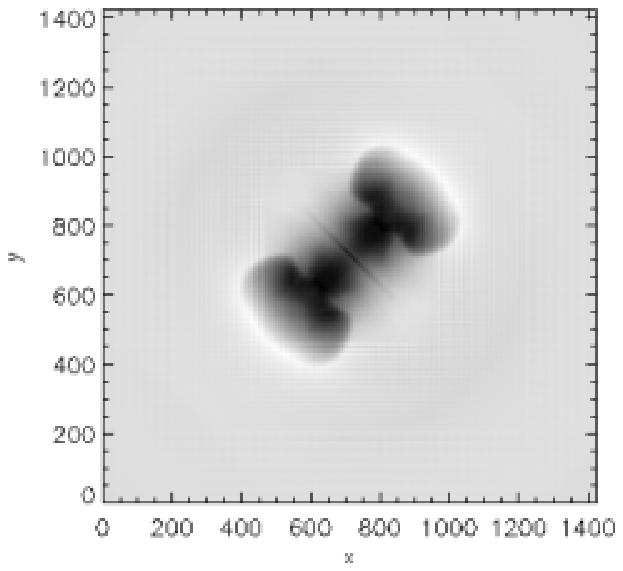}\newline
\includegraphics[trim=20 0 -10 0,clip,width=0.45\textwidth]{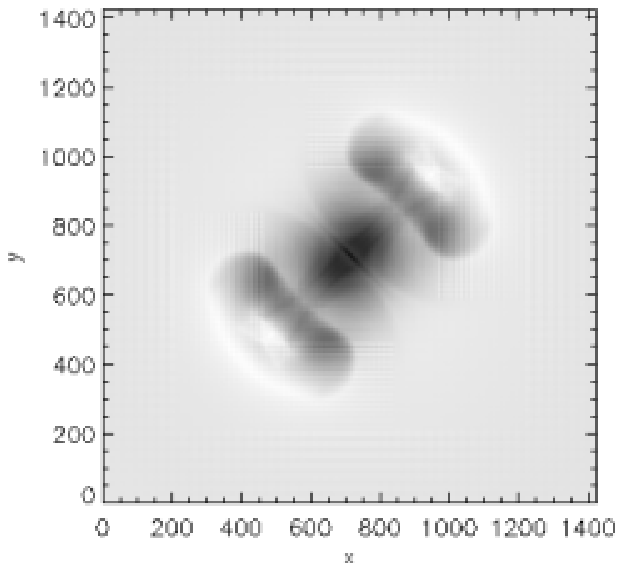}
\includegraphics[trim=20 0 -10 0,clip,width=0.45\textwidth]{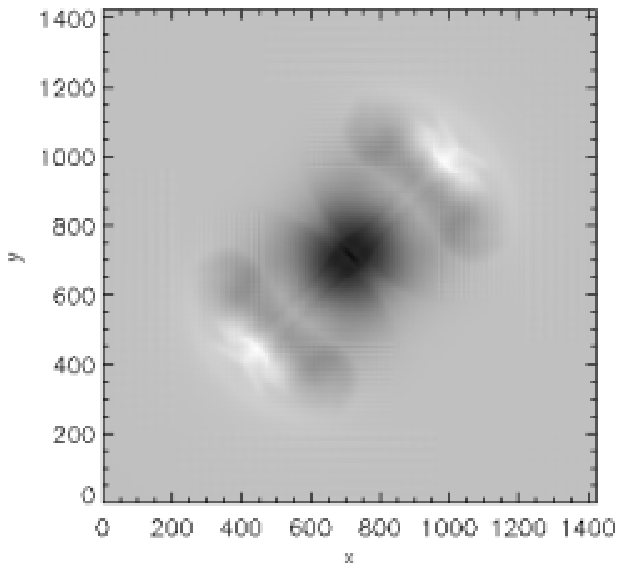}

\caption{Synthetic  {\sc Chandra} observations of a simulated galaxy cluster viewed along an axis perpendicular to the axis of the AGN, at times of 50, 100, 155 and 205 Myrs after the launch of the bubble. The first panel shows the synthetic observation. The units on the $x$ and $y$-axes are pixels. The greyscale shows the number of counts. The other panels show the X-ray flux with the image of the initial cluster subtracted, which makes it easier to discern the bubbles. }
\label{fig:simspec_z}
\end{figure*}
%FFFFFFFFFFFFFFFFFF

%FFFFFFFFFFFFFFFFFF
\begin{figure*}
\includegraphics[trim=0 0 0 0,clip,width=0.45\textwidth]{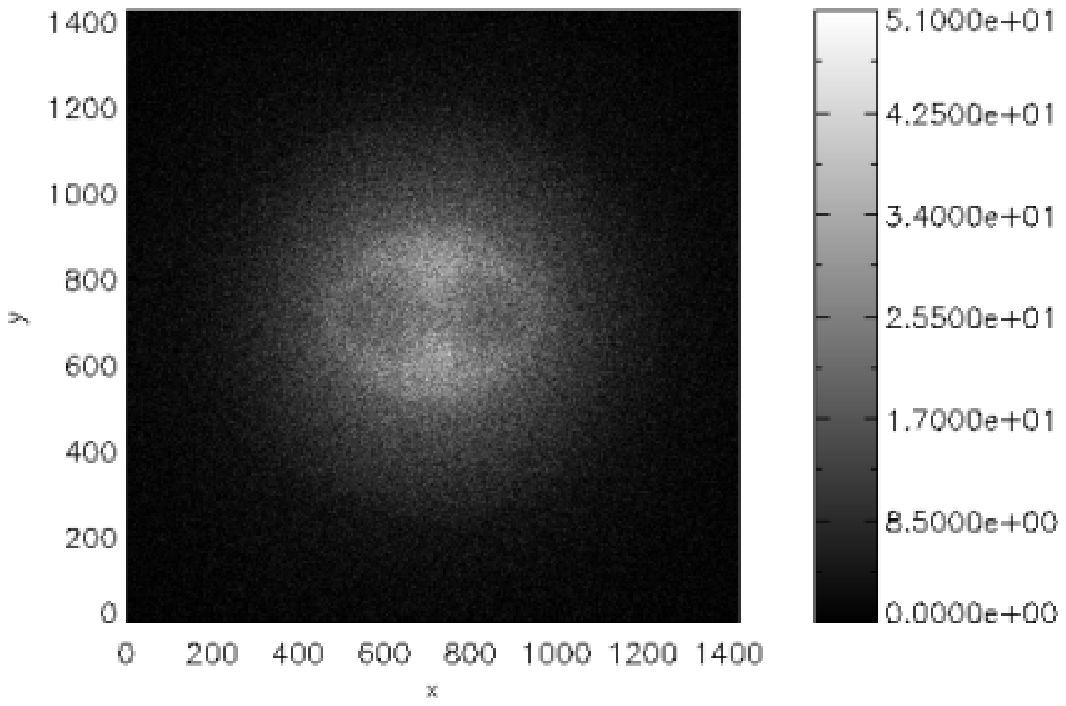}
\includegraphics[trim=0 0 0 0,clip,width=0.45\textwidth]{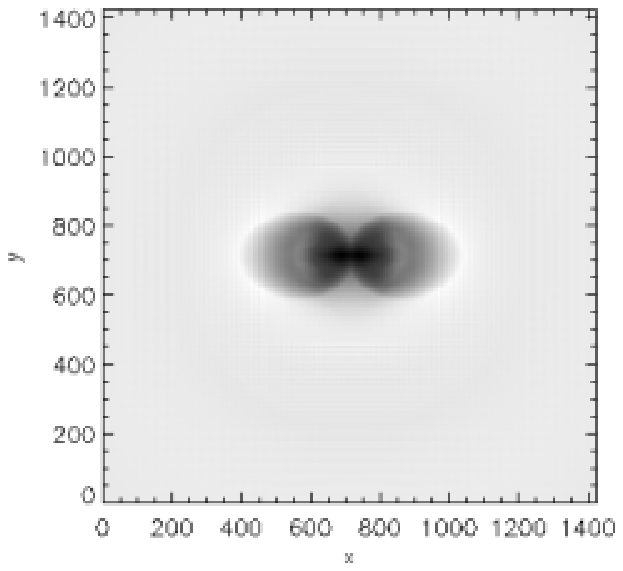}\newline
\includegraphics[trim=20 0 -10 0,clip,width=0.45\textwidth]{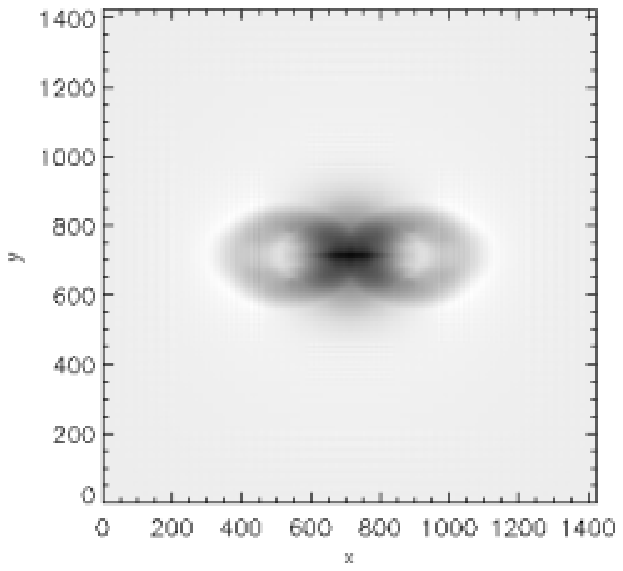}
\includegraphics[trim=20 0 -10 0,clip,width=0.45\textwidth]{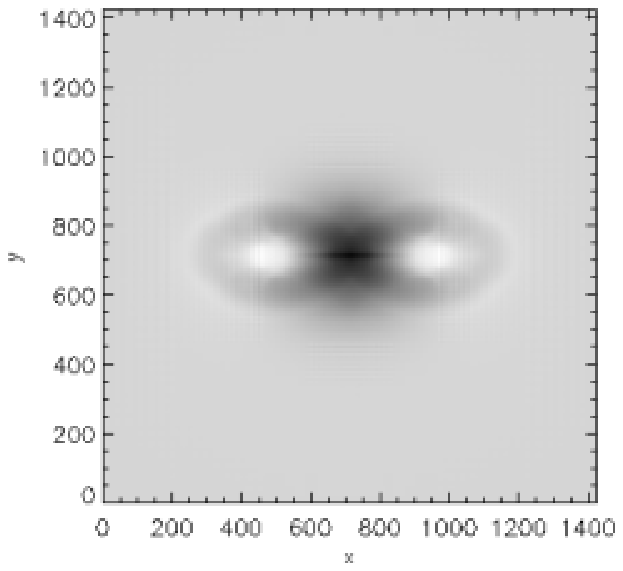}

\caption{Synthetic  {\sc Chandra} observations of a simulated galaxy cluster viewed along an axis that is inclined by 45 deg to jet axis. The panels show the count rates at times of 50, 100, 155 and 205 Myrs after the launch of the bubbles.}
\label{fig:simspec_x}
\end{figure*}
%FFFFFFFFFFFFFFFFFF

%FFFFFFFFFFFFFFFFFF
\begin{figure*}
\includegraphics[trim=0 0 0 0,clip,width=0.45\textwidth]{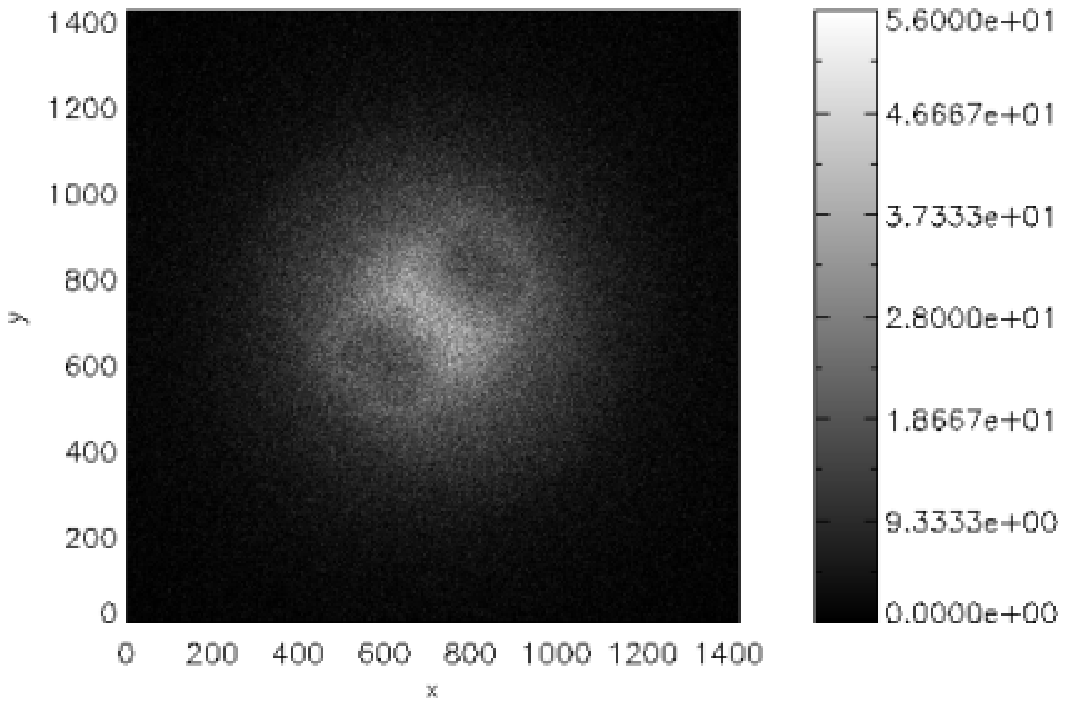}
\includegraphics[trim=0 0 0 0,clip,width=0.45\textwidth]{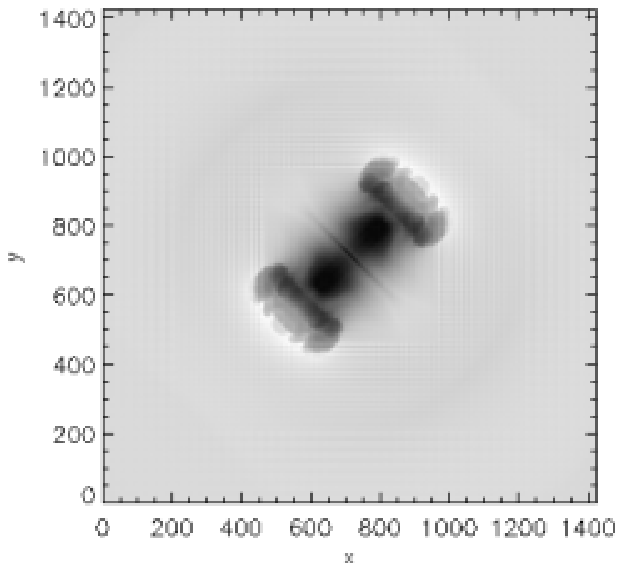}\newline
\includegraphics[trim=20 0 -10 0,clip,width=0.45\textwidth]{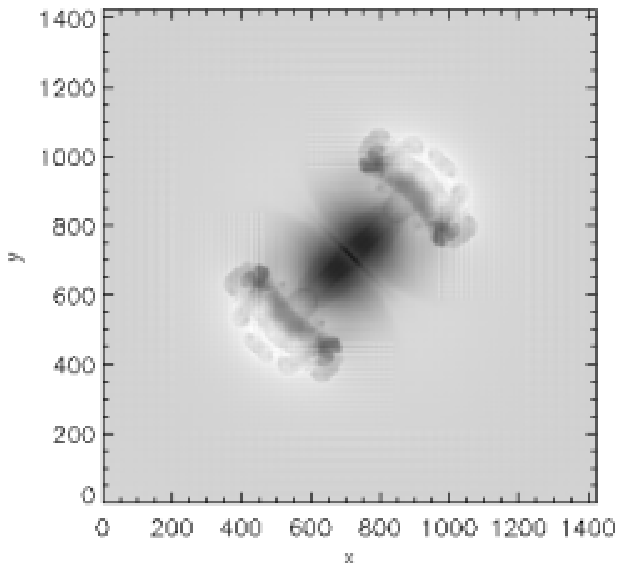}
\includegraphics[trim=20 0 -10 0,clip,width=0.45\textwidth]{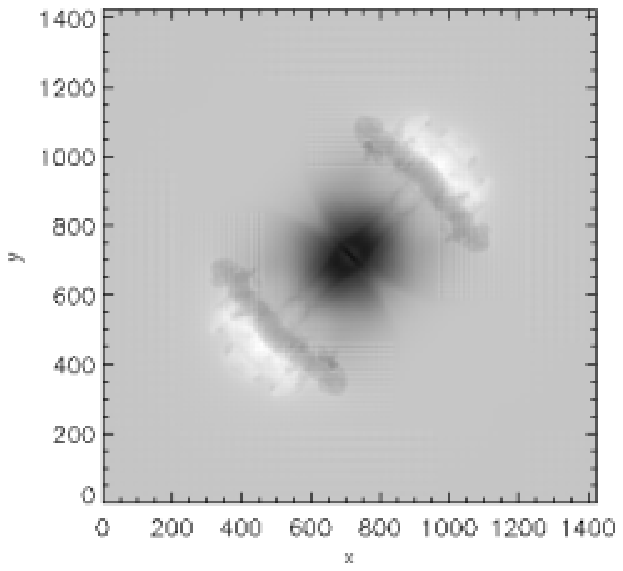}

\caption{Same as Fig.~\ref{fig:simspec_z} only without a subgrid model for turbulence.}
\label{fig:simspec_z_hydro}
\end{figure*}
%FFFFFFFFFFFFFFFFFF

%FFFFFFFFFFFFFFFFFF
\begin{figure*}
\includegraphics[trim=0 0 0 0,clip,width=0.45\textwidth]{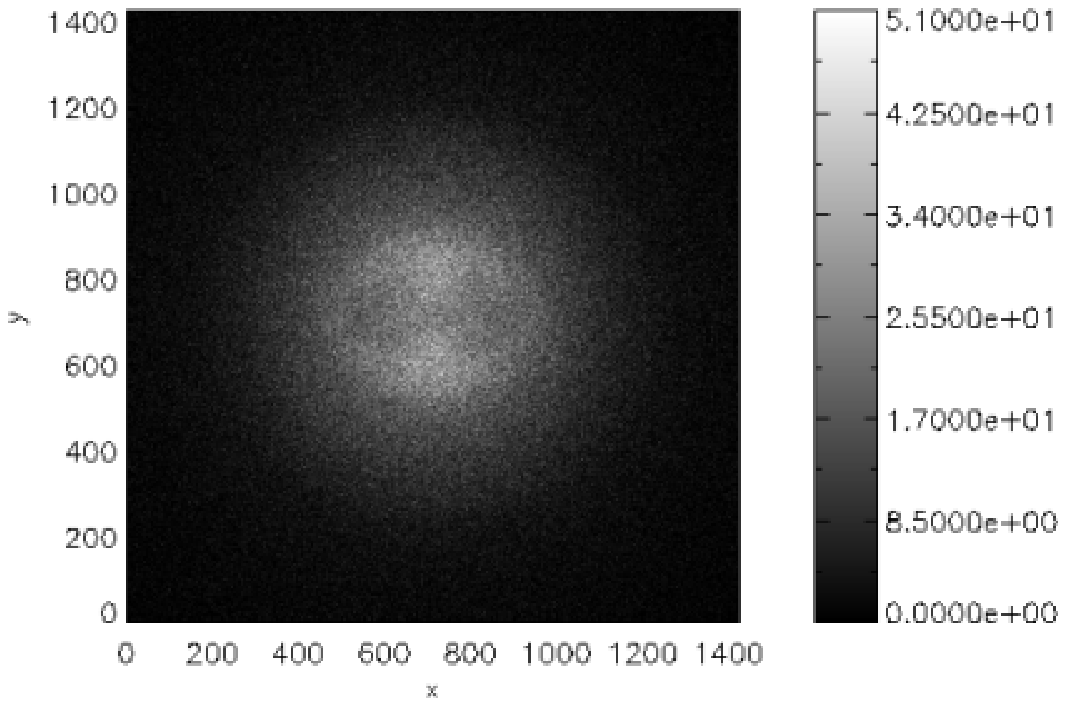}
\includegraphics[trim=0 0 0 0,clip,width=0.45\textwidth]{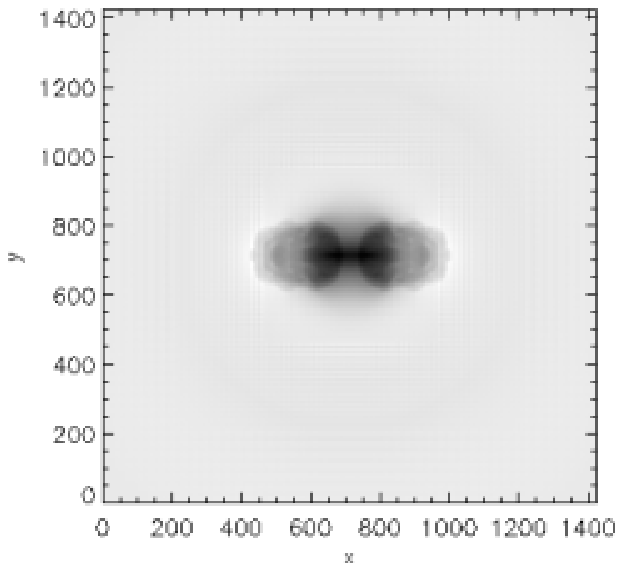}\newline
\includegraphics[trim=20 0 -10 0,clip,width=0.45\textwidth]{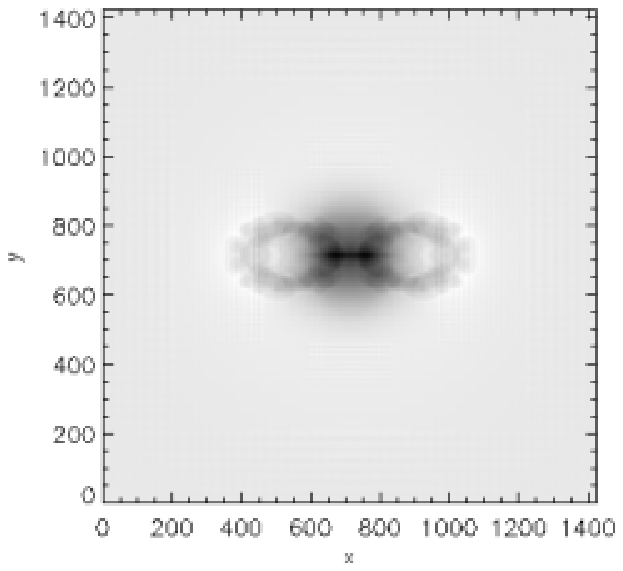}
\includegraphics[trim=20 0 -10 0,clip,width=0.45\textwidth]{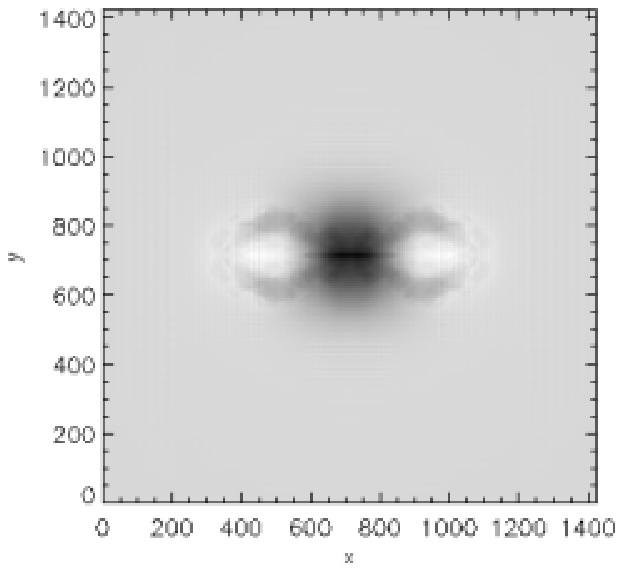}

\caption{Same as Fig.~\ref{fig:simspec_x} only without a subgrid model for turbulence. }
\label{fig:simspec_x_hydro}
\end{figure*}
%FFFFFFFFFFFFFFFFFF

%*****************
\section{Results and Discussion}
%*****************

\subsection{General Properties}

In Fig.~\ref{fig:simspec_z} - \ref{fig:simspec_x} we show
synthetic observations for the ACIS-I instrument onboard  {\sc Chandra} that
covers the band from 0.7-10 keV.  
ACIS  is a powerful tool for conducting imaging, spectroscopic and temporal studies of X-ray sources. The instrument consists of ten Charge Coupled Devices especially designed for efficient X-ray detection and spectroscopy. It is the standard instrument to study X-ray cavities.

We set the cluster to
the distance of Perseus ($z=0.017$) and the observation time to 100 ks, and first 
considered a case in which the initial bubble radii are $r_{\rm bbl} = 12$ kpc,
offset from the cluster centre by $13.5$ kpc.
Figs.~\ref{fig:simspec_z}  and \ref{fig:simspec_x}  show two different angles with respect to the axis of the
AGN. In Fig.~\ref{fig:simspec_z} the jet axis, meaning the line
connecting the bubble centres, lies in the plane of the sky. In
Fig.~\ref{fig:simspec_x}, the jet axis is inclined by 45 degrees with
respect to the line of sight.

Synthetic observations from a simulation without subgrid turbulence
corresponding to Fig.~\ref{fig:simspec_z} and Fig.~\ref{fig:simspec_x} 
are shown in Fig.~\ref{fig:simspec_z_hydro} and
Fig.~\ref{fig:simspec_x_hydro}, respectively. As apparent from the synthetic
observations, the differences produced by the subgrid turbulence are
significant, especially at times $>$ 150 Myrs. Without the subgrid
model, a bubble does not form a single coherent structure but rather looks
patchy, eventually coming apart into resolution-dependent subclumps
as described in Scannapieco \& Br\" uggen (2008).  The cavities in the run without 
subgrid turbulence also show a weaker X-ray surface brightness contrast,
even though  the ambient material gets mixed into the bubble by subgrid
turbulence. Current observations of
X-ray cavities support the picture from the subgrid simulations, and further observations of bubbles
at larger distances from cluster centres will help to probe the
stability of the bubbles and check the predictions from these kinds of
simulations.

The turbulence also affects the inferred bubble sizes, which have
been determined by subtracting a smooth, radially-symmetric background model from the
X-ray maps, as is often done with real observations and as shown in the last three panels 
in Fig.~\ref{fig:simspec_z} - \ref{fig:simspec_x_hydro}.  The corresponding radii
are shown in Fig.~\ref{fig:linradius} - \ref{fig:scaledradius}.  In
most observed cavities the diameter of all bubbles is larger than the
distance of their centres to the AGN, indicating the cavities are relatively
young.   Hence, in our simulations, we
only follow their evolution for a time up to 205 Myrs, and most of
the bubbles that have been observed should  be compared to the
first two panels shown in Fig.~\ref{fig:simspec_z} -\ref{fig:simspec_x}. 
At this stage of their evolution, all our bubbles
are found to have bright rims.  As the bubbles move outwards, they
quickly become difficult to detect as the X-ray emissivity is
proportional to the square of the density \citep{ensslin:02}. 

The decrement in surface brightness of a cavity relative to the
surrounding ICM is a strong function of its size and distance from the
cluster centre. As described in \cite{mcnamara:07}, a small bubble of
radius $R$ on the plane of the sky at a distance $r$ from the cluster
centre produces a count deficit which goes as $R^3 (1 +
r^2/r_c^2)^{-3\beta}$, where $r_c$ is the core radius of the cluster and $\beta$ a constant. 
The counts from the ICM scale as $R^2 (1 + r^2
/r_c^2)^{-3\beta +1/2}$, and the noise in this scales as its square
root. Hence, the signal-to-noise ratio scales as
$R^2(1+r^2/r_c^2)^{-3\beta/2-1/4}$, which means that cavities are
easiest to detect when they are large  and located close to the cluster centre.

\subsection{Evolution of bubble size with radius}

Diehl et al. (2008) provide simple analytic scaling relations for bubble size as a function of cluster-centric distance, which
they used to compare against their compiled observational data set.  The first is
derived from a simple $\beta$-model for the gas pressure of the ICM, given
by
\begin{equation}
  p_{\rm gas}(r)=p_0 {\left[ 1+ (r/r_c)^2 \right]^{-{3\beta\over 2}} } ,
\label{eq:peq}
\end{equation}
where the central pressure, $p_0$, is related to the central density,
$\rho_0$, and isothermal temperature, $T$, and average molecular
weight, $\bar{m}$, via the equation of state of an ideal gas with an
adiabatic index $\Gamma=5/3$.  Assuming that a bubble of initial
radius $R_{b,0}$ expands adiabatically in pressure equilibrium with
the surrounding ICM, its radius $R_{b}$ should satisfy
\begin{equation}
  p_{\rm gas}(r) R_{\rm b}^{3\Gamma} = p_{\rm gas}(r_0) R_{\rm
  b,0}^{3\Gamma} \ .
\end{equation}
Hence, the bubble radius should evolve with distance from the cluster
centre, $r$, as
\begin{eqnarray}
\label{eq:ad53}
R_{\rm b}(r)&=&
R_{b,0} \left[p_{\rm gas}(r)/p_{\rm gas}(r_0) \right]^{-\frac{1}{3 \Gamma}} \\
&=& R_{b,0} \left[ 1+ (r/r_c)^2 \right]^{{\beta\over2\Gamma}} \nonumber
\ .
\end{eqnarray}
Thus, at large radii, the cavities should grow asymptotically as
$R_{\rm  b}\propto (r/r_c)^{-{\beta/\Gamma}}$. For typical values of
$\beta=0.5$ for a cluster and $\Gamma=5/3$ this gives $R_{\rm
b}\propto (r/r_c)^{3/10}$.

Conversely, for the bubbles produced by current-dominated jets, Diehl
et al. (2008) assume a constant current along the interior and
exterior of the jet axis, such that by Amp\' ere's law $B \propto I
R^{-1}$, where $I$ is the current and $B$ is the toroidal magnetic
field strength.  This means that $p \propto B^2 \propto R^{-2}$, which
combined with Eq.\ (\ref{eq:peq}) means that   the bubble size evolves
as
\begin{eqnarray}
\label{e.rb_current}
  R_{\rm b}(r) &=&
R_{b,0} \left[p_{\rm gas}(r)/p_{\rm gas}(r_0) \right]^{-\frac{1}{2}}  \\
&=& R_{b,0} \left[ 1+ (r/r_c)^2 \right]^{{3\beta\over 4}}. \nonumber
\end{eqnarray}

As discussed above, such analytic estimates are not able to capture
effects such as mass-entrainment and distortion of the bubbles by drag
forces, which are naturally included in the simulation.  In fact, our
simulations show clearly that the cavities evolve aspherically as they
rise through the cluster, expanding quickly in the perpendicular
direction, but expanding slowly, or even becoming compressed, in the
radial direction.

\begin{figure*}
\includegraphics[trim=0 0 0 0,clip,width=0.45\textwidth]{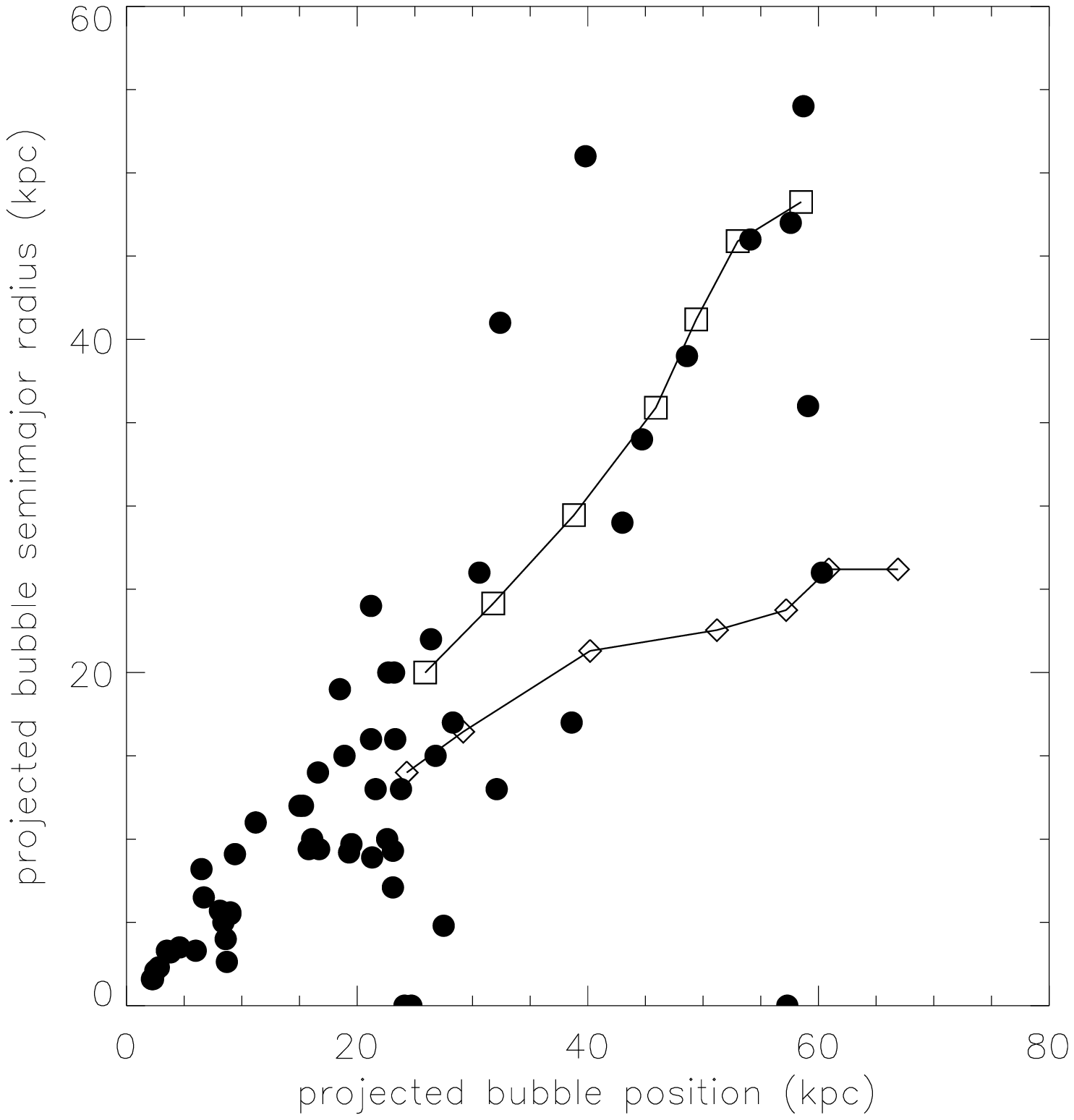}
\includegraphics[trim=0 0 0 0,clip,width=0.45\textwidth]{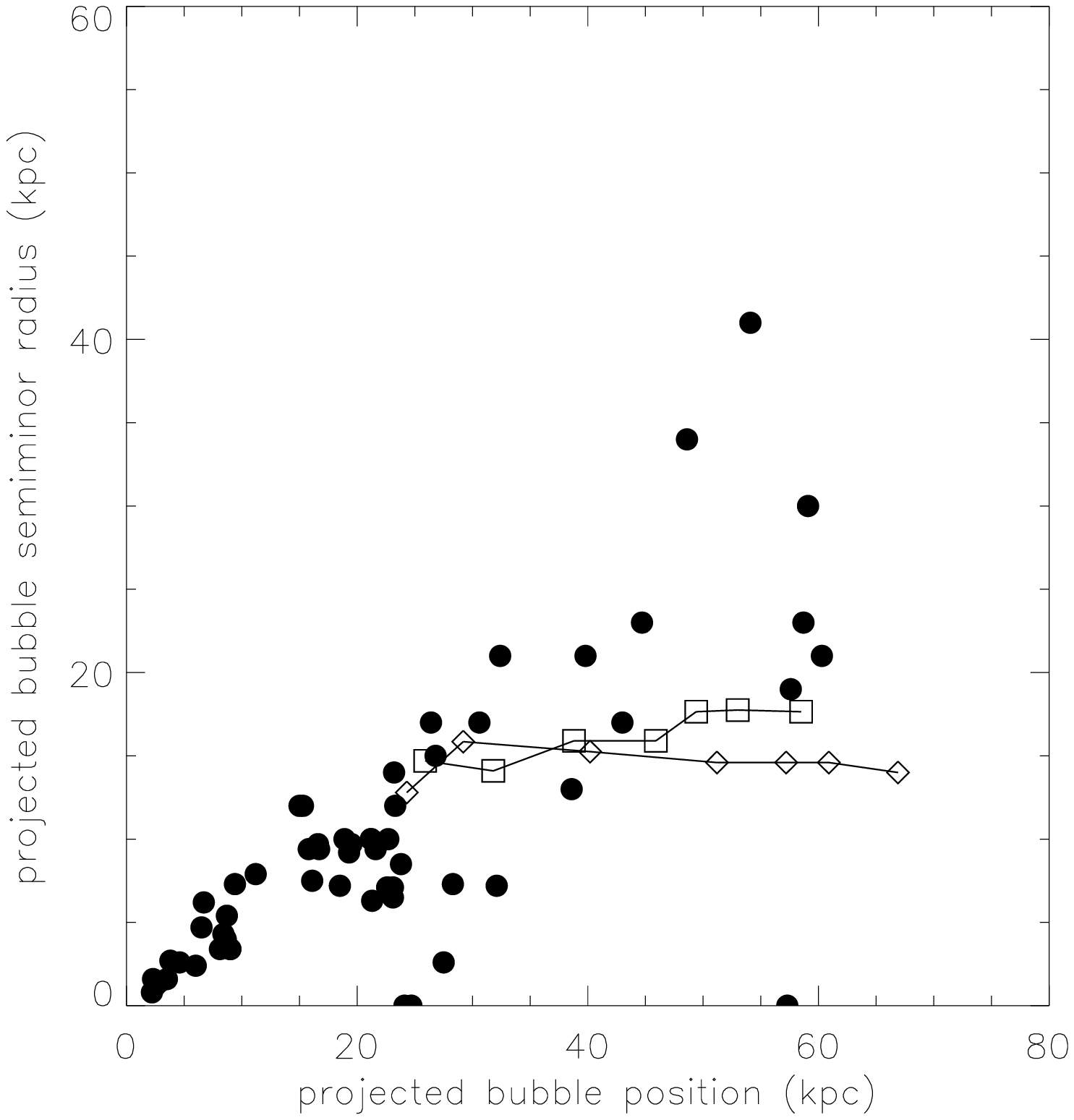}\newline
\includegraphics[trim=20 0 -10 0,clip,width=0.45\textwidth]{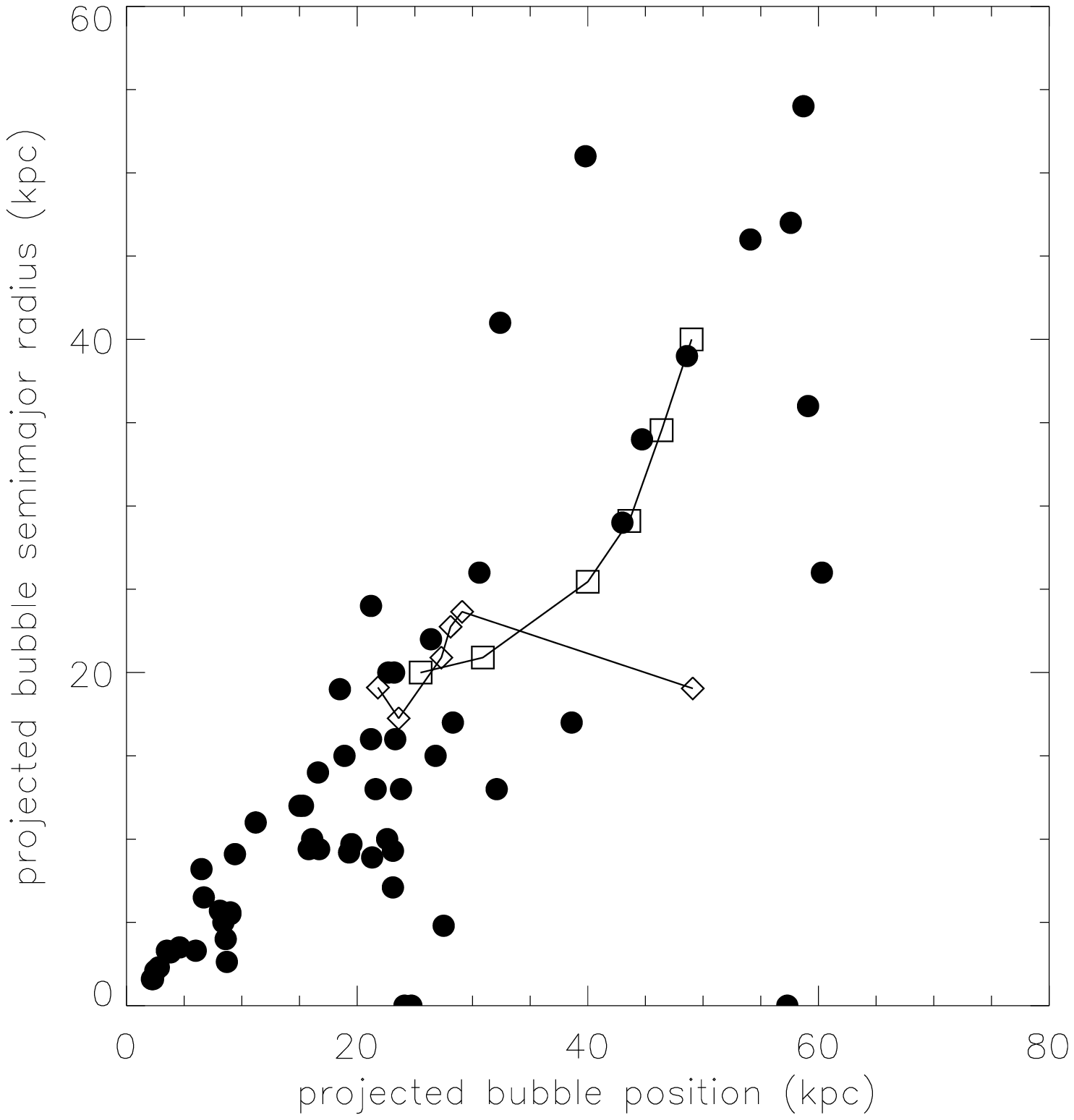}
\includegraphics[trim=20 0 -10 0,clip,width=0.45\textwidth]{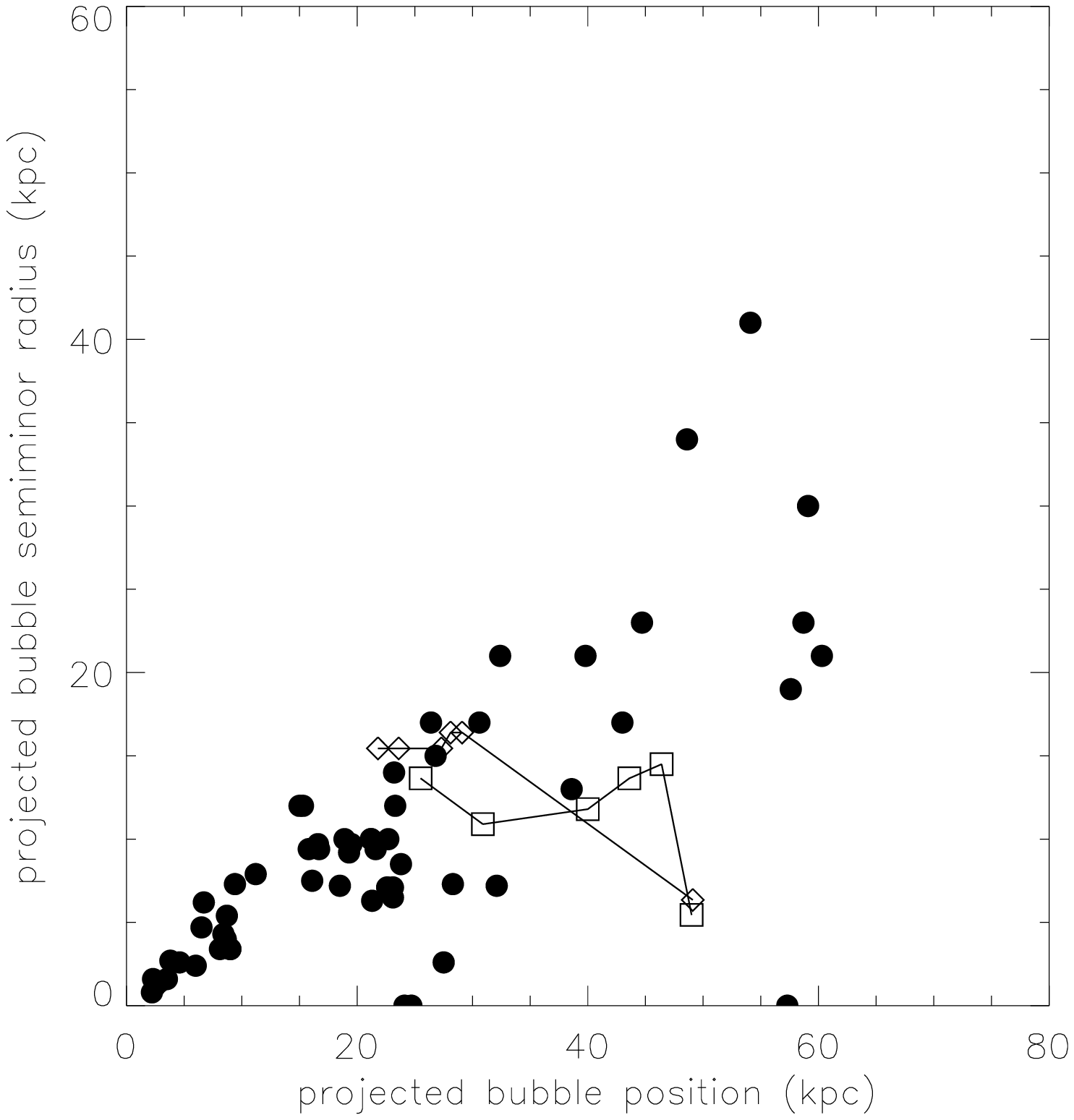}
\caption{Plot of cavity projected radii versus their projected distance from the cluster centres. The connected squares correspond to the measured radii from synthetic observations where the axis of the AGN lies in the plane of the sky. The diamonds correspond to the case where the AGN axis is inclined by 45 degrees with respect to the line of sight. The unconnected filled circles give the sizes of observed cavities compiled in Rafferty et al. (2006). The bottom row shows the results for a run without a subgrid model. {\bf Left:} semimajor radii, {\bf Right:} semiminor radii.}
\label{fig:linradius}
\end{figure*}

\begin{figure*}
\includegraphics[trim=0 0 0 0,clip,width=0.45\textwidth]{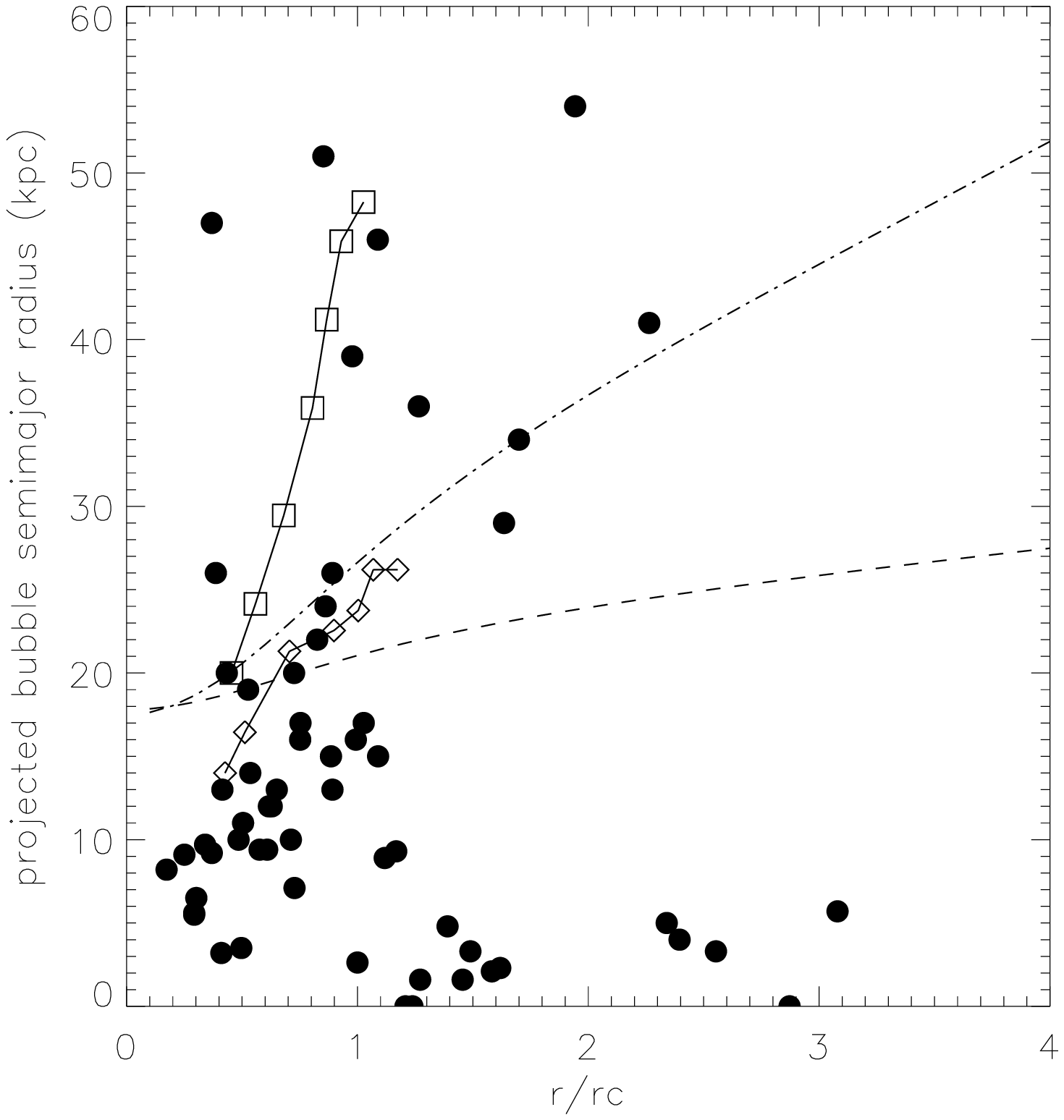}
\includegraphics[trim=0 0 0 0,clip,width=0.45\textwidth]{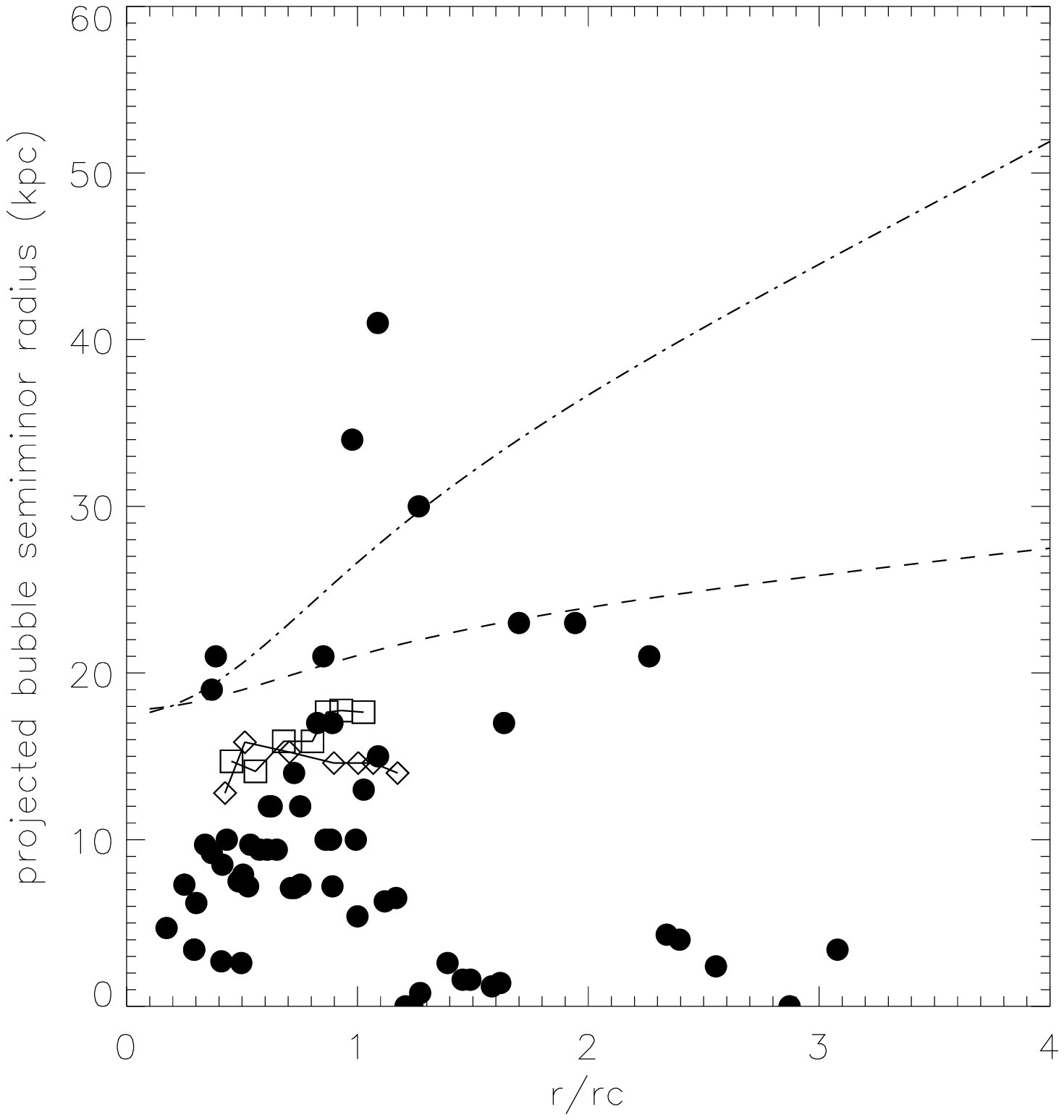}\newline
\includegraphics[trim=20 0 -10 0,clip,width=0.45\textwidth]{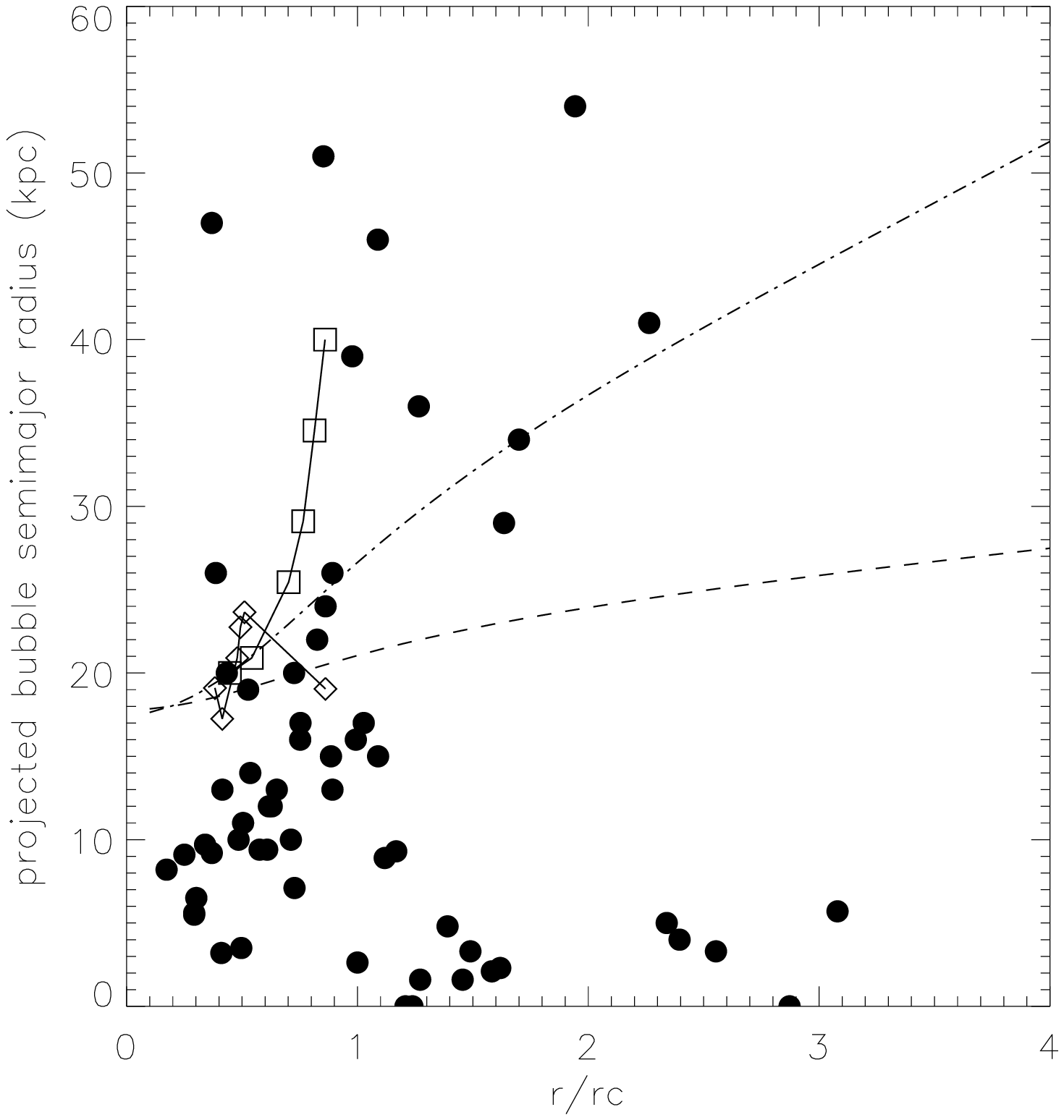}
\includegraphics[trim=20 0 -10 0,clip,width=0.45\textwidth]{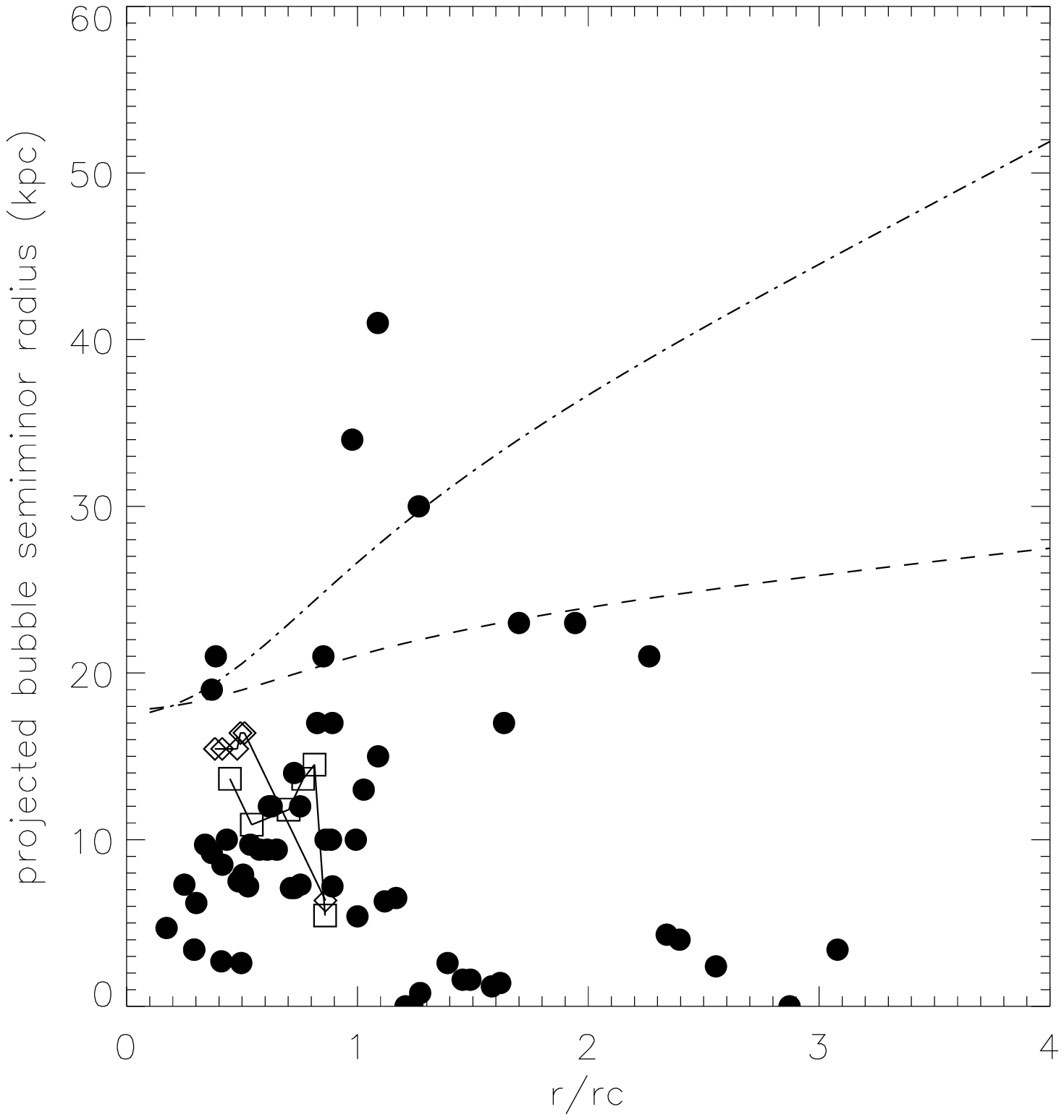}
\caption{Same as Fig.~\ref{fig:linradius}, only the bubble positions are scaled by the core radius of the cluster $r_c$. The dashed line shows the prediction by Eq.~(\ref{eq:ad53}) for a fixed value of $R_b(0)$ and $\Gamma=5/3$, and the dot-dashed line shows the prediction by Eq.~(\ref{e.rb_current}).
The unconnected filled circles give the sizes of observed cavities compiled in Rafferty et al. (2006).
{\bf Left:} semimajor radii, {\bf Right:} semiminor radii. The bottom row shows the results for a run without a subgrid model.}
\label{fig:scaledradius}
\end{figure*}

There the radial and perpendicular bubble axes evolve very differently as
shown in Figs.~\ref{fig:linradius} and \ref{fig:scaledradius}.    The radial 
extent of the bubble changes slowly as a function of
distance and time, while the perpendicular extent of the bubble
changes rapidly.   In the case where the AGN axis lies in the plane of
the sky, the semimajor axis of the surface brightness depression is
oriented perpendicular to the gravitational acceleration, and the
bubble appears to flatten as it moves outwards.  Conversely, in the
case where the AGN axis is tilted by 45 degrees with respect to the
line of sight, the projected semimajor axis lies in the projected
radial direction, and the bubble appears prolate.  In both cases, the
major axes grow much more rapidly than  the simple analytic
prescription given in Eq.\ (\ref{eq:ad53}).  On the other hand,  after
about 150 Myrs, the bubble physically flattens.  This becomes  visible
in the projection shown in Fig.~\ref{fig:simspec_x}. Note also, that
in this projection, the bubble never appears to detach from the centre
of the image for the time that we have simulated.

%FFFFFFFFFFF
%FFFFFFFFFFF

%FFFFFFFFFFF
%FFFFFFFFFFF

One can attempt to constrain bubble properties by confronting the
theoretical evolution of the bubble size with the data. However, we
do not try this here and it is difficult with the current
data. Instead, we merely point out some of the complexities associated with
such analyses, even in the non-magnetic case. In
Figs.~\ref{fig:linradius} and \ref{fig:scaledradius} we
have overplotted our measurements from synthetic observations with a
sample of bubbles by \cite{rafferty:06}. All the 64 cavities in 32
systems in that sample were observed with  {\sc Chandra}. The sample ranges
in redshift from  $z =$0.0035 to 0.545 and varies in its composition
from groups to rich clusters. For each cavity, a size and position
were measured, assuming that the cavity extends to the inner edge of
any bright surrounding emission. The projected semimajor axis, $a$, and
semiminor axis, $b$, of the cavities were measured by eye from the
exposure-corrected, unsmoothed images.

Diehl et al.\ (2008) infer that the non-magnetic models are all too
shallow to represent the data accurately, while the magnetic models
adequately reproduce the steep slope of the correlation. However, when
we examine the tangential and radial sizes separately, it is difficult
to discard a non-magnetic model. Instead, our simulations show that
the evolution of the semimajor and semiminor radii is not in obvious
contradiction to the data. In fact, the slopes that we find in
Fig.~\ref{fig:linradius} are entirely consistent with the measured
bubble sizes. In Fig.~\ref{fig:scaledradius} we have added analytic
estimates according to eqs.\ (\ref{eq:ad53}) \& (\ref{e.rb_current}),
but using the exact pressure profile for the cluster in our simulations
($p \propto n T$, with density and temperature computed from eqs.\ \ref{eq:ne} 
\& \ref{eq:te}).
The size-distance relationship of the bubbles is dramatically different
from the simple analytical estimate of Eq.\
(\ref{eq:ad53}). Especially the slope of the evolution of the
semimajor axis is much steeper than the analytical estimate for the
radius as shown by the dashed line.

In our hydrodynamic model of the cavity evolution, the effects of
projection are not trivial, and the data points are not just
systematically shifted toward smaller radii. The slopes of the
size-distance relationship of the bubbles become significantly
shallower as the AGN axis tilts toward the line of sight. The
appearance of the bubbles show stark differences, especially when the
bubbles are further than two bubble radii from the centre of the
cluster. Fig.~\ref{fig:linradius} suggest that most observed bubbles
come from systems where the AGN axis lies between 90 and 45 degrees
from the line of sight.  In fact, this is expected geometrically
for a random distribution of bubble axes.

In the case where the AGN axis lies at 45 degrees from the line of
sight, the bubbles look more like horseshoes. A confirmation of this
will require looking for bubbles at greater distances from the
centres in clusters where it is believed that the jet axis is inclined
not too far from the line of sight. This could help to verify whether
the bubbles are mainly hydrodynamic bubbles. One may speculate that
the giant cavities found in Abell 2204 resemble these late-stage
horse-shoe shaped bubbles \citep{sanders:08b}.

%FFFFFFFFFFF
\begin{figure*}
\includegraphics[trim=0 0 0 0,clip,width=0.45\textwidth]{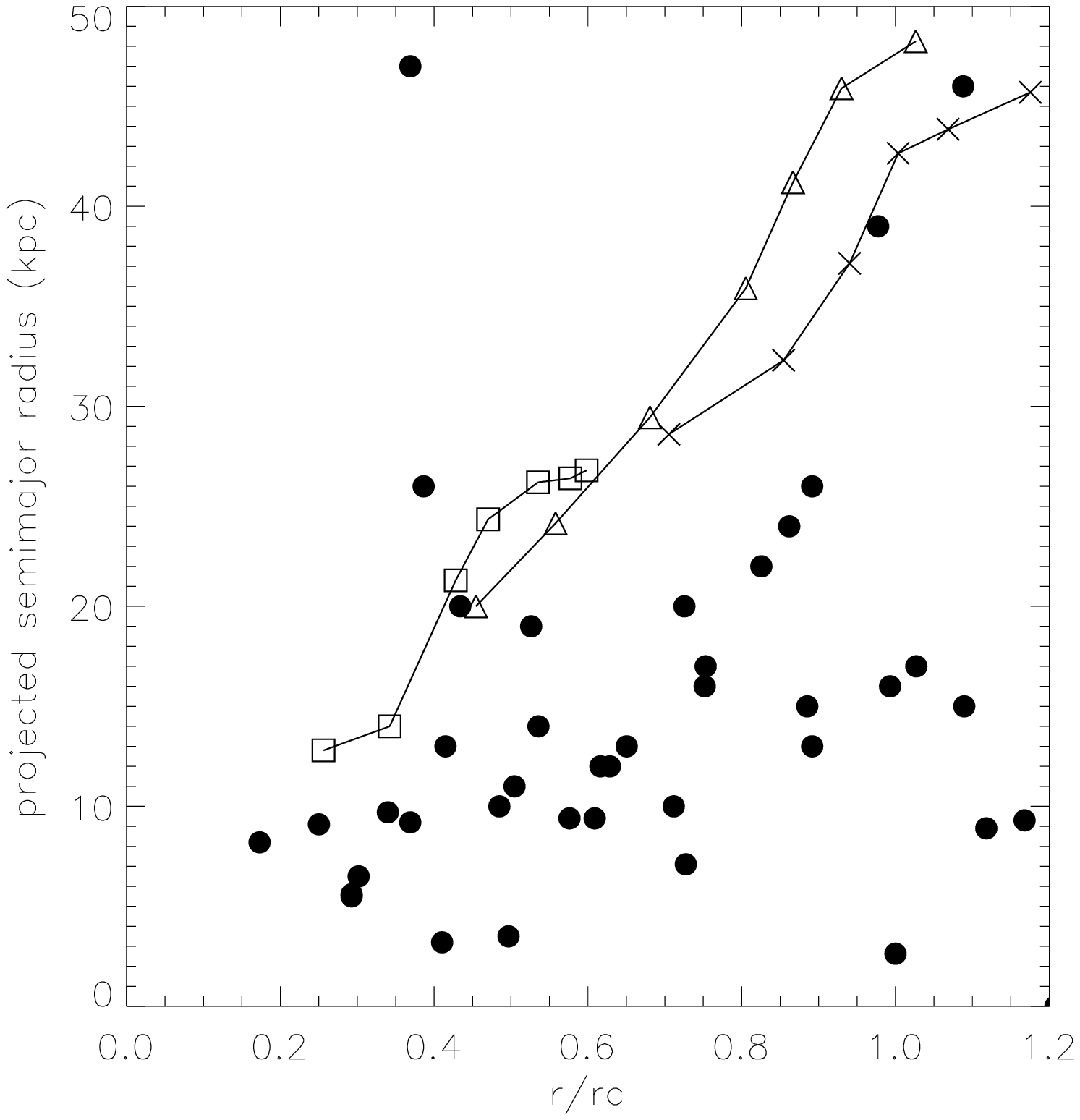}
\includegraphics[trim=0 0 0 0,clip,width=0.45\textwidth]{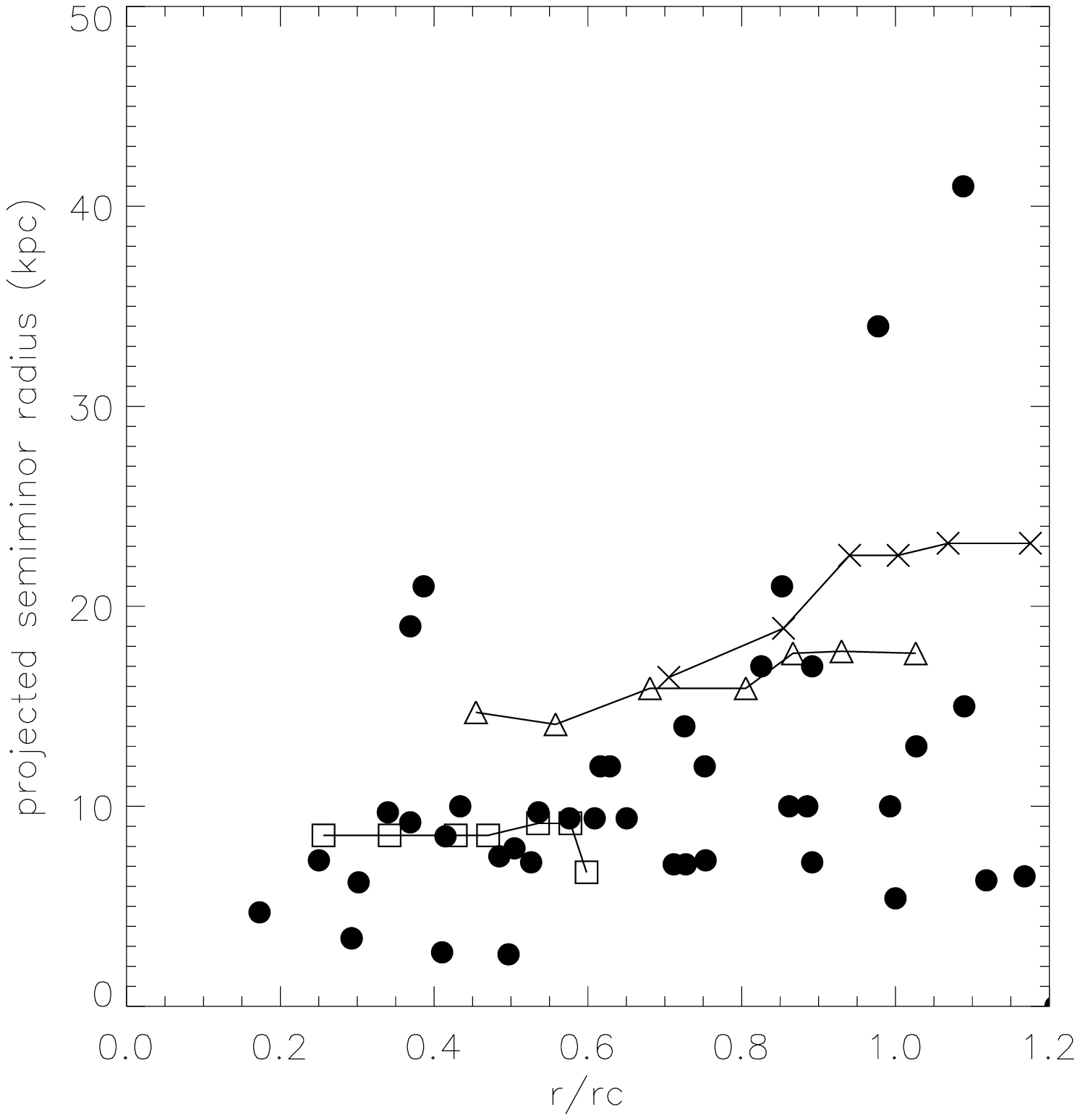}\newline
\includegraphics[trim=20 0 -10 0,clip,width=0.45\textwidth]{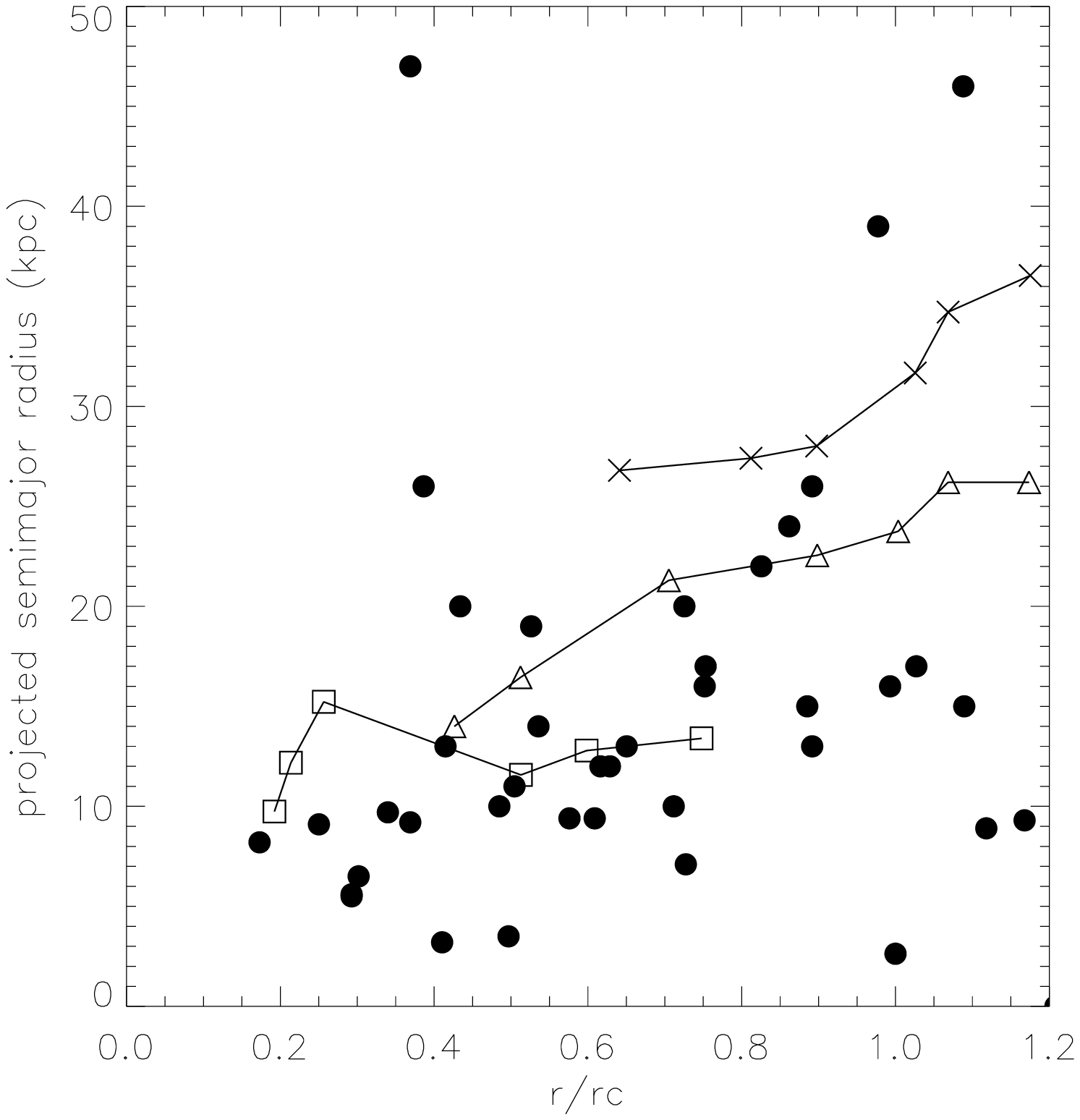}
\includegraphics[trim=20 0 -10 0,clip,width=0.45\textwidth]{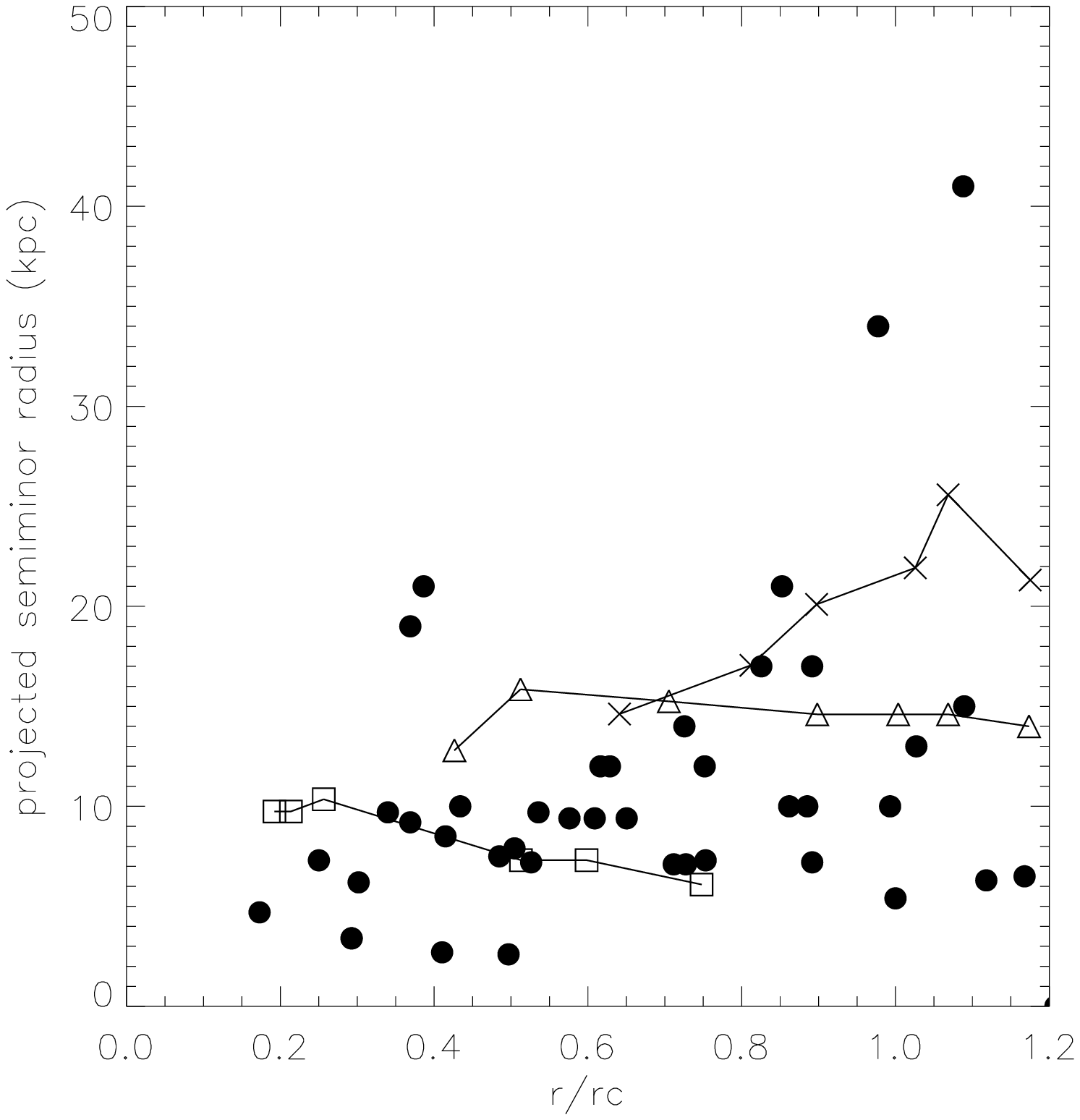}

\caption{Semimajor (left) and semiminor (right) projected radii versus bubble positions scaled by the core radius of the cluster $r_c$. The lines correspond to runs with three different sizes for the initial bubbles. Squares: small bubbles (6 kpc), triangles: medium-sized bubbles (12 kpc), crosses: large bubbles (18 kpc). The unconnected filled circles give the sizes of observed cavities compiled in Rafferty et al. (2006). The top row corresponds to the case where the axis of the AGN lies in the plane of the sky; the bottom row displays results for the case where the AGN axis is inclined by 45 degrees with respect to the line of sight.}
\label{fig:sizes}
\end{figure*} 
%FFFFFFFFFFF

The spread in bubble radii at a given distance from the centre in
Fig.~\ref{fig:linradius} - \ref{fig:scaledradius} is due to a spread
in the initial bubble size (or energy) which is determined by
properties of the AGN and the host cluster. Diehl et al. (2008) find
that the mass of the supermassive black hole and the core radius account for most of the
spread.  To study the impact of these differences further, we carried
out two additional runs, both of which included subgrid turbulence.
In one run, we reduced the initial radius of the bubbles to $r_{\rm
bbl} = 6$ kpc, and offset them from the centre by $8.5$ kpc, and in a
second run we increased the initial radius of the bubbles to $r_{\rm
bbl} = 18$ kpc, offset from the centre by $19.5$ kpc.  The results of
these runs are shown in Fig. \ref{fig:sizes}. 

In general, the evolution of the minor and major radii are similar across
runs.  If the AGN axis lies in the plane of the sky,
the major axis, which is perpendicular to the 
radial direction, evolves much more rapidly than expected from analytic
estimates for all choices of initial bubble size.
Likewise, in this projection, regardless of initial bubble size, the minor axis grows 
only very slowly as the cavity moves outward.  In all cases, the differences between
these runs are well within the observational scatter.
In the projection where the AGN axis is inclined 
by 45 degrees with respect to the line of sight,  the major axis, which 
is now parallel to the radial direction, grows at a moderate rate in all three runs.
In all runs this evolution is much more gradual than measured when the bubbles
are viewed in the plane of the sky.  In all runs, the minor axis of the bubbles
remain roughly constant in this projection, even occasionally shrinking
slightly. Thus, initial cavity size and radial offset seem the be relatively minor issues
as compared with the anisotropic evolution of the cavities and the angle from
which this evolution is viewed.

Finally, Fig.~\ref{fig:pvplot} shows the $pV$ energy as function of
projected distance from the cluster centre. {As discussed above, the $pV$
energy inferred from observations appears to increase with
increasing distance, even though this trend is less pronounced at small energies. This may in part be due to the fact that bubbles
tend to be overpressured close to their origin, and expand to reach
pressure equilibrium as they rise though the ICM. However, excess
pressure in the observed cavities is difficult to measure, except
indirectly through the presence of shocks and sound waves.

A second possibility for the radial increase in inferred $pV$ is the
entrainment of ambient material.  In fact, this is the main reason for
the increase in our simulations, as in our model the bubbles quickly
reach pressure equilibrium with the surrounding medium,  well before
moving noticeably out from the cluster centre.  Over the 200 Myrs that
we simulate, we find that the inferred $pV$ energy for the same bubble
grows by a factor of  2 in the case where the AGN axis lies in the
plane of the sky.  In the case where the AGN axis is inclined by 45
degrees with respect to the line of sight, the inferred $pV$ energy
grows hardly at all.  In the latter case, the greater apparent pressures (because the bubbles are further in) are offset by the bubbles appearing to be smaller.

The measured energies are more
reliable when the bubbles are further from the centre of the cluster,
which unfortunately, is where they are most difficult to observe.
Overall, the error introduced into estimates of the $pV$ energy by
these effects is of the same magnitude as the error related to the
unknown equation of state of the plasma inside the bubbles.

\section{Conclusions}

The nature of AGN-driven X-ray cavities remains one of the major
outstanding  questions in understanding the physics of cool-core
galaxy  clusters. While the  observed morphologies and sizes of the
cavities provide us with useful clues  as to the processes at work in
these regions, interpretation of the observations is far from
straightforward.

In this study we have focused on a model in which X-ray cavities
evolve purely hydrodynamically, and reconstructed their detailed
evolution using of two major tools:  AMR simulations that include
subgrid turbulence and synthetic X-ray software that produces realistic
observations from these simulations.  Together these tools allow us to
capture such important effects as mass-entrainment, distortion of the
bubbles by drag forces, and observational effects. These effects
lead to an evolution that is  drastically different than expected from
simple analytic estimates.

In particular, we find that while the radial extent of the cavities
changes slowly as  a function of distance and time, they expand
rapidly in the perpendicular  direction. The result is a complex
evolution that is highly dependent on viewing angle and difficult to
compare conclusively with observations.  Although analytic
estimates of non-magnetic models evolve too slowly to match the observations,
our simulations show that the evolution of the
semimajor and semiminor radii is not in obvious contradiction to the
data.  In fact, our simulations naturally reproduce the overall 
trend for inferred $pV$ energies of observed X-ray cavities to increase as 
a function of distance from the cluster centre, an effect that is largely
due to mixing of entrained material into the rising cavities.
The size evolution we find is in general good agreement
with the measured bubble sizes, although there is a large spread in
observational data, and a strong dependence on the direction from
which the cavities are viewed.  Indeed, the flattening and projection
of X-ray cavities have much stronger effects on our results than
initial bubble size and radius.

Finally, some words of caution. Our simulations start from a very
simplified setup. In reality, the bubbles are inflated by jets
continuously over some time.  Also the host system will be dynamic,
leading to a more complex evolution than the simulations presented
here can capture. One example are the bubbles in Hydra that do not
look like the bubbles we have shown here.  In the Hydra cluster, a rapid
succession of outbursts from the AGN have continued to fuel the
bubbles to produce a partially interconnected series of bubbles that
still have some momentum imparted by the original jet. More realistic
simulations of specific systems that are designed to reproduce all
observed features are needed in order to make better inferences on the
nature of X-ray cavities. This, in combination with low-frequency
radio observations, will help to understand these elusive and crucial
features of cosmic structure.

%FFFFFFFFFFF
\begin{figure}
\centering\resizebox{0.8\hsize}{!}{%
\includegraphics{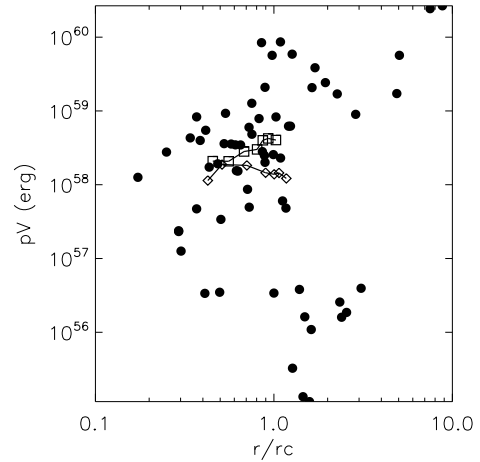}}
\caption{$pV$ energy as function of projected distance from the cluster centre. Again squares correspond to synthetic observations where the axis of the AGN lies in the plane of the sky; diamonds correspond to the case where the AGN axis is inclined by 45 degrees with respect to the line of sight.
The unconnected filled circles give the sizes of observed cavities compiled in Rafferty et al. (2006).}
\label{fig:pvplot}
\end{figure} 
%FFFFFFFFFFF

%*******************************************************************
%************ A C K N O W L E D G E M E N T S **********************

\section*{Acknowledgements}

MB acknowledges the support by the DFG grant BR 2026/3 within the Priority
Programme ``Witnesses of Cosmic History'' and the supercomputing grants NIC
2195 and 2256 at the John-Neumann Institut at the Forschungszentrum J\"ulich.
All simulations were conducted on the ÒSaguaroÓ cluster operated by the 
Fulton School of Engineering at Arizona State University.
The results presented were produced using the FLASH code, a product of the DOE
ASC/Alliances-funded Center for Astrophysical Thermonuclear Flashes at the
University of Chicago.

%*******************************************************************
%************ A P P E N D I X  **********************
%\appendix

%*******************************************************************
%*************** R E F E R E N C E S *******************************
%*******************************************************************
%
\bibliographystyle{mn2e}
\bibliography{%
BIBLIOGRAPHY/icm_conditions,%
BIBLIOGRAPHY/bubble_evol%
}

\bsp

\label{lastpage}

\end{document}